\documentclass[twocolumn,prd,superscriptaddress,altaffilletter,nofootinbib]{revtex4-1}
\usepackage{float}
\usepackage{amsmath}
\usepackage{amssymb}
\usepackage{amsfonts}
\usepackage{comment}
\usepackage{graphicx}
\usepackage{tabularx}
\usepackage{hyperref}
\usepackage[nolist]{acronym}
\usepackage[usenames,dvipsnames]{xcolor}
\usepackage[utf8]{inputenc}
\usepackage{xspace}
\usepackage{tikz}
\usepackage{siunitx}
\usetikzlibrary{arrows,shapes,trees,decorations.pathreplacing}
\allowdisplaybreaks

\DeclareSIUnit \year{yr}
\DeclareSIUnit \erg{erg}
\DeclareSIUnit \parsec{pc}
\DeclareSIUnit \sol{M_{\odot}}

\begin{document}

\title{A gravitational-wave limit on the Chandrasekhar mass of dark matter}

\author{Divya Singh}
\affiliation{Department of Physics, The Pennsylvania State University, University Park, PA, 16802, USA}
\affiliation{Institute for Gravitation and the Cosmos, The Pennsylvania State University, University Park, PA 16802, USA}

\author{Michael Ryan}
\affiliation{Department of Physics, The Pennsylvania State University, University Park, PA, 16802, USA}
\affiliation{Institute for Gravitation and the Cosmos, The Pennsylvania State University, University Park, PA 16802, USA}

\author{Ryan Magee}
\affiliation{Department of Physics, The Pennsylvania State University, University Park, PA, 16802, USA}
\affiliation{Institute for Gravitation and the Cosmos, The Pennsylvania State University, University Park, PA 16802, USA}

\author{Towsifa Akhter}
\affiliation{Department of Physics, The Pennsylvania State University, University Park, PA, 16802, USA}

\author{Sarah Shandera}
\affiliation{Department of Physics, The Pennsylvania State University, University Park, PA, 16802, USA}
\affiliation{Institute for Gravitation and the Cosmos, The Pennsylvania State University, University Park, PA 16802, USA}

\author{Donghui Jeong}
\affiliation{Institute for Gravitation and the Cosmos, The Pennsylvania State University, University Park, PA 16802, USA}
\affiliation{Department of Astronomy and Astrophysics, The Pennsylvania State University, University Park, PA 16802, USA}

\author{Chad Hanna}
\affiliation{Department of Physics, The Pennsylvania State University, University Park, PA, 16802, USA}
\affiliation{Department of Astronomy and Astrophysics, The Pennsylvania State University, University Park, PA 16802, USA}
\affiliation{Institute for Gravitation and the Cosmos, The Pennsylvania State University, University Park, PA 16802, USA}
\affiliation{Institute for CyberScience, The Pennsylvania State University, University Park, PA 16802, USA}

\date{\today}
\begin{abstract}

We explore a new paradigm to study dissipative dark matter models using gravitational-wave observations. We consider a dark atomic model which predicts the formation of binary black holes such as GW190425 while obeying constraints from large-scale structure, and improving on the missing satellite problem. Using LIGO and Virgo gravitational-wave data from 12th September 2015 to 1st October 2019, we show that interpreting GW190425 as a dark matter black-hole binary limits the Chandrasekhar mass for dark matter to be below \SI{1.4}{\sol} at $> 99.9\%$ confidence implying that the dark proton is heavier than 0.95 GeV, while also suggesting that the molecular energy-level spacing of dark molecules lies near $10^{-3}$ \si{\electronvolt} and constraining the cooling rate of dark matter at low temperatures. 

\end{abstract}

\maketitle

\section{Introduction}\label{sec:intro}

The Universe provides a remarkable laboratory where matter reaches temperatures and densities inaccessible on Earth. 
Consequently, phenomena observed by astrophysicists have been the first source of precision measurements of fundamental properties of atoms and molecules~\cite{Hoyle:1954zz, refId0}, and have uncovered new particle physics~\cite{doi:10.1146/annurev.aa.27.090189.003213}. 
But as rich as the study of light-emitting astrophysical objects is, it reaches only $15\%$ of the matter in the Universe. 
The rest of the matter is dark, and its particle nature is unknown. 
While experiments have slowly expanded their reach into the enormous parameter space of possible dark matter particle interaction channels and strengths~\cite{Penning_2018,Dainese:2019rgk,PhysRevD.101.052002,PhysRevLett.118.221803,Diamond:2019luc,Roszkowski_2018,Conrad:2017pms}, they have yet to produce a detection. 
However, gravitational-wave observatories like LIGO~\cite{TheLIGOScientific:2014jea} and Virgo~\cite{TheVirgo:2014hva} provide a new way of detecting astrophysical objects where dark matter is on the same footing as visible matter.
Gravitational-wave detections do not require any coupling between dark and visible matter; they only require that dark matter gravitates.

The connection between dark matter and gravitational-wave sources has been explored previously, perhaps most notably with the first detection of a binary black-hole (BBH), GW150914~\cite{Abbott:2016blz}, by the LIGO and Virgo collaborations (LVC). 
Primordial black-holes(PBHs)~\cite{Carr2016}, long theorized to be a component of dark matter, were invoked as a possible progenitor~\cite{Bird:2016dcv, Sasaki:2016jop} due to the previously unobserved high mass of GW150914's constituents. 
Other gravitational-wave studies~\cite{Clesse:2020ghq, Abbott:2005pf,PhysRevLett.121.231103,PhysRevLett.123.161102,PhysRevD.101.104041} have explored the PBH paradigm for dark matter and set upper limits on the fraction of dark matter in PBHs. 

While PBHs do not invoke the existence of new particles in the universe today, their presence requires yet undiscovered inflationary-era physics. 
Here we present a new paradigm which connects gravitational-wave observations to dark matter through dissipative dark matter. 
In contrast to PBHs, dissipative dark matter works within the standard cosmological history, and introduces dark particle physics with a complexity resembling that of the Standard Model. 
Dissipative dark matter models allow for the formation of compact objects through the cooling and eventual gravitational collapse of dark matter.
Therefore, gravitational-wave observations of BBH mergers could either confirm or severely constrain dissipative dark matter models. 
We note that this paradigm is in contrast with the weakly-interacting massive particle (WIMP) paradigm that includes a single cosmologically relevant dark matter particle whose primary interaction is with the Standard Model. 
The WIMP paradigm is appealing in its simplicity, but it is increasingly under pressure from both the lack of results in direct detection experiments~\cite{Bertone:2018krk} and from astrophysical data that may prefer self-interacting dark matter~\cite{Bullock:2017xww,Essig:2018pzq,Choquette:2018lvq,Latif:2018kqv,DAmico:2017lqj,2020Univ....6..107D}. 

In this article, we argue that gravitational-wave observations may already be probing the properties of dark matter. 
Of particular interest to our work is GW190425~\cite{Abbott:2020uma}, which was a $5\sigma$ deviation in total mass from all other observed Galactic binary neutron stars (BNS) prompting speculation that the signal could be from a BBH, a neutron star - black hole (NSBH), or a PBH~\cite{Clesse:2020ghq} merger. 
Here, we show that if GW190425 is caused by compact objects made of atomic dark matter, it constrains the dark proton mass directly through the Chandrasekhar Limit, and implies cooling properties of dark matter which could provide a possible parameter space for dark matter particle interactions.

\section{\label{sec:DBHform}Black Hole Formation from dissipative dark matter}

Black holes are simple objects in general relativity, carrying only three parameters: mass, spin, and electric charge~\cite{Kerr:1963ud}. 
The Chandrasekhar limit~\cite{Chandrasekhar:1931ih} provides a fundamental upper mass limit for degenerate stars where Fermi degeneracy pressure balances the gravitational force toward collapse. In the absence of nuclear forces, any compact object heavier than the Chandrasekhar limit must be a black hole. While the Chandrasekhar limit gives the lowest possible mass for a black hole from gravitational collapse, it makes no further prediction about the actual black hole mass spectrum, which depends on the formation process of black holes. 
In scenarios where black holes made of dark matter form via astrophysical processes, the spectrum of black holes may be used to constrain the properties of dark matter, if a dark analogue to the Chandrasekhar limit exists.

 We consider here a dark matter model with simple chemistry, forming only a single atom analogous to hydrogen. 
All dark black holes (DBHs) are then formed by direct collapse. 
In a model of dark matter with an abundant spin-1/2 particle of mass $m_x$, analogous to the standard model proton (mass $m_p$), the dark matter Chandrasekhar mass limit ($M_{\textrm{DC}}$) is $\SI{1.4}{\sol} (m_p/m_x)^2$.
While dark matter may well have more complex chemistry, including nuclear forces only yields marginal changes to the model given that both the Chandrasekhar limit and the maximum neutron star mass are of the order $(\hbar c / G)^{3/2}/m_p^2 = \SI{1.9}{\sol}$. 

Atomic dark matter~\cite{Feng_2009} consists of two spin-1/2 particles oppositely charged under a new force. 
The dark fermion masses are $m_x$ and $m_c$, with $m_x > m_c$. The particles interact through a force analogous to electromagnetism, mediated by a massless dark photon, with interaction strength determined by the dark fine structure constant $\alpha_D$ (where $\alpha = 1/137$ is the fine structure constant in the Standard Model). 
The oppositely charged particles can form bound states similar to atomic and molecular hydrogen, $H_D$ and $H_{D,2}$, with the lighter dark particle analogous to the standard model electron. 
In regions of sufficient density~\cite{PhysRevLett.120.051102}, the dark matter gas may cool by standard radiative cooling processes including recombination, Bremsstrahlung, and collisional excitation of the atoms or molecules~\cite{PhysRevD.96.123001}. 
In lower-density regions, cooling is inefficient, which is consistent with the observed dark matter halos. 
Even if dark matter is dissipative, only a fraction of the gas will cool, and of that, only a fraction will end up in compact objects.
We call the fraction of dark matter that ultimately ends up in DBHs, $f$.

We use gravitational-wave observations to constrain $f$, by modeling the population of binary DBHs after Population-III stars. 
Briefly, Population-III stars form from midsize halos dominated by atomic hydrogen. Radiative processes cool the gas, allowing it to collapse inwards.   Hydrogen molecules are formed as the gas contracts, enabling additional cooling. At sufficient density, the primary cooling processes become inefficient and the gas begins reheating, with some of the gas eventually fragmenting into small protostars\cite{Glover2012}. Assuming the same general process can occur for dissipative dark matter halos, some of the dark matter will collapse into a pseudo-protostar, but, lacking pressure from nuclear fusion-induced radiation, will collapse further into a black hole. The mass spectrum of dark matter compact objects formed this way is obtainable, to a first approximation, by rescaling results from Population-III star formation literature~\cite{1976MNRAS.176..483R,1976MNRAS.176..367L,Abel:2001pr,doi:10.1146/annurev.astro.42.053102.134034}. The minimum Jeans mass in the dark matter gas can be computed analytically and depends on microphysical parameters. Simulations of Pop III star formation follow the hydrodynamical evolution of the gas and find that the smallest proto-stars formed are significantly larger than the analytic minimum. Combining the minimum Jeans mass calculation for atomic dark matter with a rescaling from simulation results gives the minimum mass of dark black holes~\cite{Shandera:2018xkn},
\begin{align}
\label{eq:Mdj}
	M_{\textrm{DJ}} \approx \SI{800}{\sol} \left(\frac{m_p}{m_x}\right)^{5/2} \left(\frac{m_c}{m_e}\right)^{1/2} \left(\frac{\alpha_D}{\alpha}\right)^{1/2}.
\end{align}
\linebreak
This minimum mass depends on the coldest temperature the gas can reach, proportional to the energy gap of the lowest rotational modes of the $H_{2}$ molecule for Population-III stars and $H_{D,2}$ for the dark compact objects. 

LVC data directly constrains the dark Chandrasekhar limit, since any black holes observed must be larger than $M_{DC}$. However, if the black holes are formed by direct collapse in a cooling gas, then $M_{\textrm{DJ}}$ can also be constrained. While simulations of dissipative dark matter are required to provide the precise numerical coefficient in Eq.(\ref{eq:Mdj}), this interpretation is a powerful one as it allows the molecular energy gap to be inferred. We present both constraints below in order to demonstrate how the spectrum of compact objects can provide new constraints on the particle physics and chemistry of the dark sector. 

We use Population-III star formation studies~\cite{doi:10.1146/annurev.astro.42.053102.134034} to set several parameters or parameter ranges of the dark black holes that should be determined by hydrodynamics and gravity rather than by particle physics. We take the initial mass function to be given by $\mathcal{P}(m) \propto m^{-b}$, ranging from $M_{\textrm{min}}$ up to $M_\textrm{max} =  r M_{\textrm{min}}$ for some $m_x$, $m_c$ and $\alpha_D$. The range of $r$ extends from 2 to 1000 distributed log-uniformly, while $b$ ranges from -1 to 2 uniformly in our study. We denote the DBH population parameters $\{M_{\textrm{min}}, r , b\}$ as a vector $\overline{\theta}$. The merger time for binaries is given by the Peters formula~\cite{PhysRev.131.435}, which depends on the initial distribution of the eccentricities and the semi-major axis of the binary. Again based on Population-III star populations~\cite{doi:10.1146/annurev.astro.42.053102.134034, Hartwig:2016nde}, we draw the eccentricities of the binaries, $e$, from $\mathcal{P}(e) \propto e^{m}$ with $0.1 < e < 1$ and $m = 1$. The semi-major axis, $a$, follows the distribution $\mathcal{P}(x) \propto x^k$ where $x = \log_{10}(a/a*)$, $k = -1/2$ with the range of $a$ proportional to $M^{1/3}$, and we rescale $a*$. The mass ratio, $q = m_\textrm{light} / m_\textrm{heavy}$ follows the distribution $\mathcal{P}(q) \propto q^n$ with $n = -0.55$. The merger time is taken to be  \SI{10}{\giga\year}, appropriate for binaries formed at high redshifts ($z > 1$). We use LVC data to constrain $f$, assuming the fraction of DBHs in binaries is $f_\textrm{binary} = 0.26$~\cite{doi:10.1146/annurev.astro.42.053102.134034}.

\section{Constraining properties of dark black hole populations}\label{sec:paramConstraints}

Gravitational-wave observations from the LVC~\cite{LIGOScientific:2018mvr} infer the component masses $m_1$ and $m_2$ of merging black holes. Specifically the binary chirp mass $\mathcal{M} = (m_1 m_2)^{0.6}/ (m_1 + m_2)^{0.2}$ is measured very accurately. Given LVC observations as a function of chirp mass, we constrain $f$ and $M_{\textrm{min}}$ by calculating the posterior probability of $f$ and $M_{\textrm{min}}$ conditioned on the predicted DBH merger event rate $R$, and the sensitivity of the LIGO and Virgo detectors~\cite{Abadie:2010cg}. We use a Bayesian approach to infer $f$ and $\overline{\theta} = \{M_{\textrm{min}}, r , b\}$, $P(f, \overline{\theta} | \overline{\mu}) \propto P(\overline{\mu} | f, \overline{\theta}) P(f, \overline{\theta})$, where $\overline{\mu}$ is a vector corresponding to the expected number of DBH detections from the LVC in pre-defined chirp mass bins over $\mathcal{M}\in$ [\SI{0.2}, \SI{200}{\sol}], such that $\mu_i$ is the mean event count in the $i^{th}$ chirp mass interval. $P(f, \overline{\theta})$ denotes the prior probabilities of $f, M_{\textrm{min}}, r$ and $b$ assumed to be independent of each other. We use uniform priors for $f$, $M_{\textrm{min}}$ and $b$, and a logarithmic prior for $r$. In order to obtain the posterior probability distribution of $f$ and $\overline{\theta}$, we must predict the probability of obtaining event counts, $n_i$ in a bin with $\mu_i = R_i(f, \overline{\theta}) V_i(\overline{\theta})T$, where $R_i$ is the predicted rate of mergers for DBH binaries, $V_i$ is the spatial volume surveyed by a gravitational-wave detector, and $T$  is the observation time.

The DBH merger rates are modeled as a function of chirp mass, $\mathcal{M}$ and $f$ following ~\cite{Shandera:2018xkn}, and computed in the $i^{th}$ chirp mass interval using
\begin{align}
\begin{split}\label{eq:rate}
	R_i (\mathcal{M} = m_i  | f, \overline{\theta}) = 
	P_i (m_i | t_{\textrm{m}}, \overline{\theta})
	\times
	\left(\frac{dP(t_{\textrm {m}} = \SI{10}{\giga\year}  | \overline{\theta})}{dt}\right) \\
	\times
	\left(\frac{\rho_{\textrm{DM}} \times f \times f_{\textrm{binary}}}
	{\langle M \rangle}\right),
\end{split}
\end{align}
where  $\rho_{\textrm{DM}}$ = \SI{3.3e19}{\sol\per\giga\parsec\cubed} is the density of dark matter in the universe, $f_{\textrm{binary}} = 0.26$, and $P(\mathcal{M} | t_m, \overline{\theta})$ is the chirp mass distribution of binary systems that would merge at a given merger time $t_{\textrm{m}}$, $P(t_{\textrm{m}}=\SI{10}{\giga\year} | \overline{\theta})$ is the probability that the merger time of the binary is \SI{10}{\giga\year}, roughly the age of the universe, and $\langle M \rangle$ is the mean component mass of DBH given the initial mass distribution, computed for some $\overline{\theta}$. 

The LVC has detected several binary black hole mergers during the first, second and third advanced LIGO-Virgo observing runs. We include all published events from these observing runs~\cite{LIGOScientific:2018mvr,Abbott:2020niy} excluding the known BNS detection, GW170817~\cite{TheLIGOScientific:2017qsa}, as definitely not a DBH detection. We assume that the event count $n_i$ observed by the LVC in a given chirp mass interval $i$ follows a Poisson distribution~\cite{Biswas:2007ni}, $\mathcal{P}(\mu_i | f, \overline{\theta}) = \mu_i^{n_i}e^{-\mu_i}/n_i !$, with independent $\mu_i$. 

To approximate $V_iT$ in our predetermined chirp mass bins, we use the horizon distance which depends on the chirp mass as $\mathcal{M}^{5/6}$~\cite{Abadie:2010cg}. For a given chirp mass, we use the predicted distribution of mass ratios for DBH binaries to compute a weighted average over all possible binaries allowed for that chirp mass. Equation~\ref{eq:VT1} was used to compute the $VT$ for a compact binary uniquely described by its mass ratio, $q$ and chirp mass, $m_i$. We then use equation~\ref{eq:VT2} to calculate the weighted average of the $VT$s for a given chirp mass over mass ranges representative of the search parameter spaces used for compact binary coalescences conducted in data from the second and third observing runs of LVC~\cite{Mukherjee:2018yra}. The posterior distribution for $f$ and $M_{\textrm{min}}$ is finally obtained by marginalizing the four dimensional posterior distribution over the undetermined parameters, $r$ and $b$. More information on methods is provided in appendices~\ref{appendix:posteriors} and \ref{appendix:VT}.

\begin{figure*}
  \includegraphics[width=\textwidth]{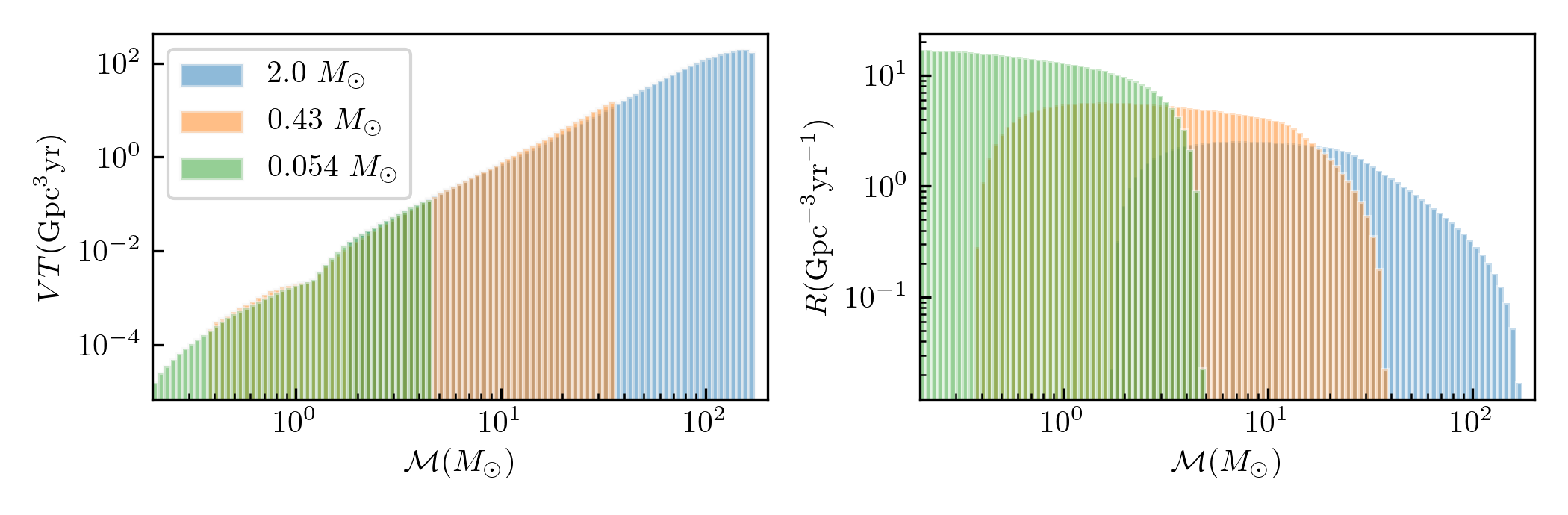}
  \caption{Sensitive volume (\textit{left}) and rate of DBH mergers (\textit{right}) evaluated over pre-defined chirp mass bins for different values of $M_{\textrm{min}}$. The slope of the initial mass distribution, $b = 2$ and ratio, $r = 100$ for both figures, while the fraction of dark matter in DBHs, $f$ is set to be $10^{-4}$. }
  \label{fig:VT_R}
\end{figure*}

\begin{figure}[ht]
  \includegraphics[width=\columnwidth]{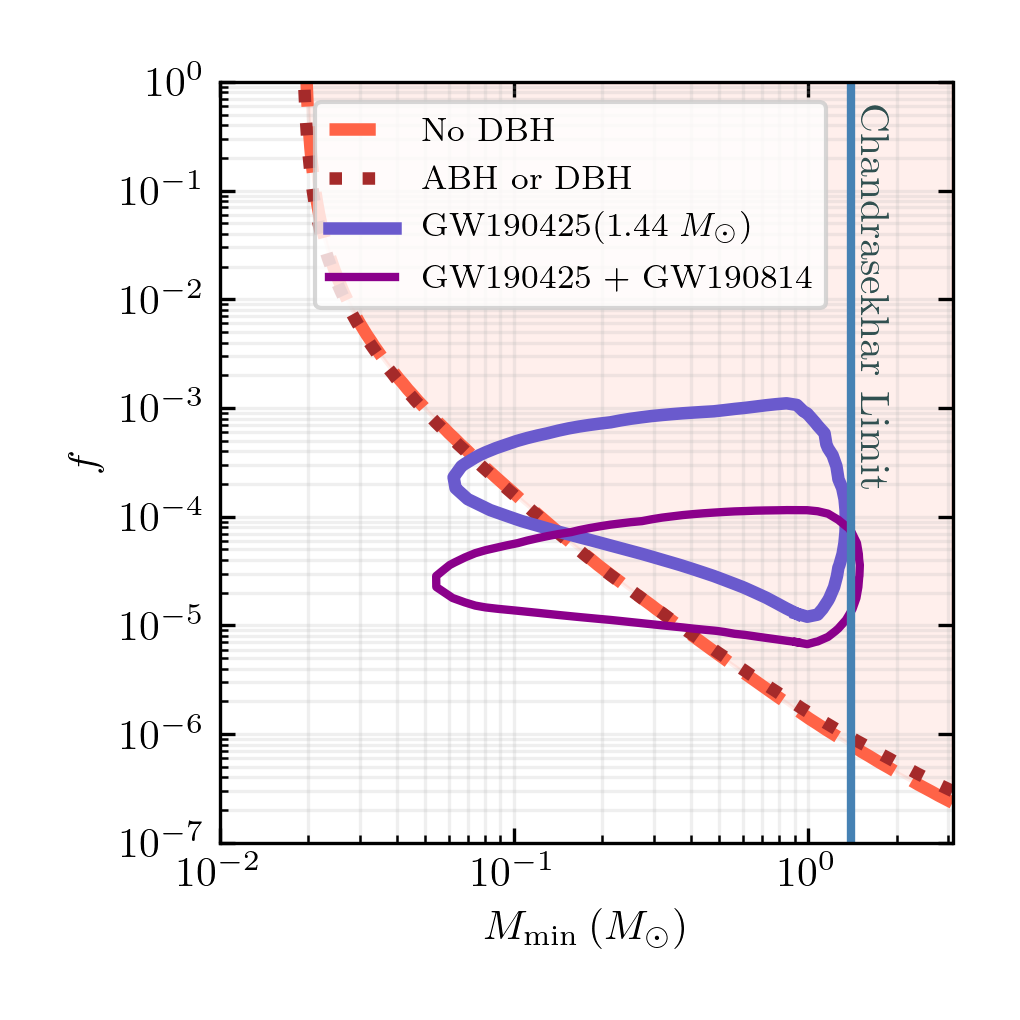}
  \caption{Constraint on the fraction of dark matter $f$ in DBHs against the minimum allowed mass of DBH $M_{\textrm{min}}$ for the dissipative dark matter model of~\cite{Shandera:2018xkn} if (a) none of LVC's binary black holes are DBHs (\textit{dashed}), (b) all LVC detections are either from astrophysical or DBH binaries with equal probability (agnostic about origin) (\textit{dotted}), (c) GW190425 is a DBH binary detection (\textit{blue}), and (d) GW190425 and GW190814 are from DBH binaries (\textit{pink}). Constraints for (c) and (d) are computed under an agnostic assumption about the origins of other LVC events. The Chandrasekhar limit for conventional black holes (\SI{1.4}{\sol}) is plotted for reference. The contours mark 90\% confidence regions.}
\label{fig:constraints}
\end{figure}

Assuming GW190425~\cite{Abbott:2020uma}, is a binary DBH signal and that the other LVC detected events from the first, second and the first half of the third observing runs~\cite{LIGOScientific:2018mvr,Abbott:2020niy} are either conventional BBH or DBH binaries with equal probabilities, we report the constraints on the parameters $f (10^{-10} - 1)$, and $M_{\textrm{min}}(10^{-3} - \SI{3}{\sol})$ in figure \ref{fig:constraints}. 
The chirp mass $1.4^{-0.02}_{+0.02} \si{\sol}$ and total mass $(M = m_1+ m_2)$ $3.4^{-0.2}_{+0.3} \si{\sol}$ of this system are five standard deviations greater than those observed for known Galactic BNS systems, implying that GW190425 could possibly be from a BBH or an NSBH merger. 
However, if one of the components is a black-hole, it would be the smallest known black hole by a significant margin, and such small black holes have not been observed prior to LIGO~\cite{2011ApJ...741..103F}, though they have been proposed to form through various channels~\cite{Faber:2012rw, Gupta:2019nwj, Bramante:2017ulk, Belczynski_2012}. 
Another detection, GW190814~\cite{Abbott:2020khf} with chirp mass $6.1^{+0.06}_{-0.05}\si{\sol}$, has prompted speculation about possible dark matter origins for the binary~\cite{Clesse:2020ghq}.
Figure \ref{fig:constraints} shows that the addition of GW190814 as a DBH detection widens the constraint on $M_{\textrm{min}}$ while constraining $f$ to a narrower range of lower values. 
The lighter mass event dominates the constraints on $M_{\textrm min}$ when we have a wide range of allowed masses.
It is evident from equation~\ref{eq:rate} is that we have more DBHs at lower masses than higher masses. This factor later cancels with the sensitive volume, where the horizon distance scales as $\mathcal{M}^{5/6}$, such that $f$ is lowered when we include the heavier binary.


\begin{table}
\begin{tabularx}{\columnwidth}[t]{p{1.35in}p{.6in}p{.7in}p{.65in}}
\footnotesize{\textbf{Observed DBH binary}} & \footnotesize{$\mathbf{\mathcal{M}}$\textbf{/ \si{\sol}}} & \footnotesize{$\mathbf{M_{\textrm{min}}}$\textbf{/ \si{\sol}}} & \footnotesize{$\mathbf{m_x}$\textbf{/GeV}}\\
\hline
 GW190425 & 1.44 & 0.062--1.34 & 0.95--4.44 \\ 
 GW190425, GW190814 & 1.44, 6.1 &0.054--1.50 & 0.91--4.76 \\

\end{tabularx}
\caption{Probable minimum masses of dark black-holes, $M_{\textrm{min}}$, and the corresponding heavy fermion masses, $m_x$ for the two cases of observed dark black-hole binaries. Heavy fermion masses are computed using the dark matter Chandrasekhar mass limit, by setting $M_{DC}=M_{\textrm min}$ determined from the data.}
\label{table:masses}
\end{table}

If GW190425 is a DBH binary, $M_{\textrm{min}}$ lies below \SI{1.4}{\sol}, the Chandrasekhar limit for astrophysical black holes, to greater than $99.9\%$ confidence. Using the Chandrasekhar limit equation, we get dark heavy fermion masses ($m_x$) ranging from 0.95 -- \SI{4.44}{\giga\electronvolt} for the possible values of $M_{\textrm{min}}$ between 0.062 -- 1.34 \si{\sol}, which is greater than the mass of the proton $(m_p)$, \SI{0.938 }{\giga\electronvolt}. We report the possible heavy fermion masses from the Chandrasekhar mass limit for the considered DBH detections in table~\ref{table:masses}.

Alternatively, using Population-III star formation as a guide to black hole formation in the atomic dark matter scenario~\cite{Shandera:2018xkn}, we find that dark-hydrogen molecular cooling dissipates the average kinetic energy to \SI{2.2}{\milli\electronvolt} for $M_{\textrm{DJ}} = \SI{0.92}{\sol}$ ( roughly the most probable value assuming only  GW190425 is a DBH), assuming $m_x = \SI{14}{\giga\electronvolt}$, $m_c = \SI{325}{\kilo\electronvolt}$, and $\alpha_D = 0.01$. The Chandrasekhar mass for $m_x = \SI{14}{\giga\electronvolt}$ is much lower, at \SI{6e-3}{\sol}. 

If none of the compact binary coalescences observed by the LVC are DBH binaries, we get upper limits on $f$ for a range of possible $M_{\textrm{min}}$, also shown in figure \ref{fig:constraints}. For $M_{\textrm{min}}$ lower than \SI{0.2e-2}{\sol}, $f$ is completely unconstrained. This is due to the inconclusive limits on $f$ for low mass compact object binaries from searches for sub-solar mass binaries in gravitational-wave data~\cite{PhysRevLett.121.231103, PhysRevLett.123.161102, PhysRevD.101.104041}, and partly due to the assumed priors for our parameters, specifically on $r=M_{\textrm{max}}/M_{\textrm{min}}$.

\section{Constraints on cooling rates}

Besides the lower bound of the dark-matter particle mass from the Chandrasekhar limit, the existence of DBHs also provides a novel way to limit the energy dissipation rate of dark matter. 
Forming a DBH requires the originating dark matter halo to lose order one of its kinetic energy density $\mathcal{E}_{\rm k}$, on a time-scale shorter than its free-fall time $t_{\rm ff}$, which is the time for the halo to collapse under its own gravity. The cooling rate $\Lambda$ in the birthplace halos of DBHs must satisfy \cite{PhysRevLett.120.051102}, 
\begin{align}
\begin{split}
	\Lambda \geq \left(\frac{\mathcal{E}_{\rm k}}{t_{\rm ff}}\right) \left(\frac{1}{n_{\rm D}}\right)^2,
\end{split}\label{eq:lambda_ff}
\end{align}
where the normalizing factor $n_{\rm D}$ is the number density of the dominant mass component of dissipative dark matter, analogous to the number density of hydrogen, $n_{\rm H}$ (as opposed to the total number density, $n$) in the Standard Model. Note that, while the free-fall time only depends on $n_{\rm D}$, the kinetic energy density represents all dark-matter particles. That difference introduces a non-trivial temperature dependence to equation (\ref{eq:lambda_ff}) as the ionization fraction varies with temperature.

\begin{figure}
\includegraphics[width=\columnwidth]{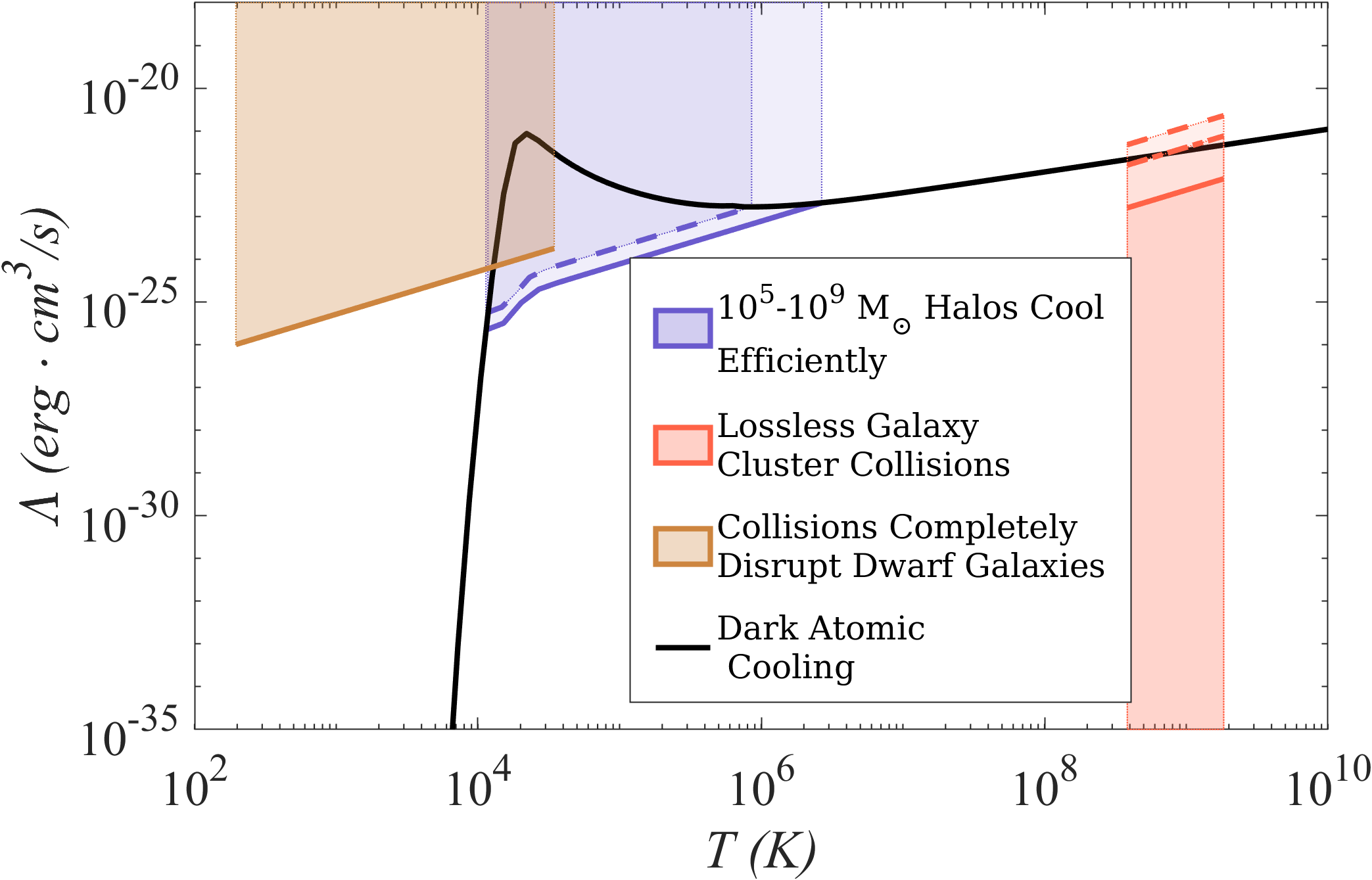}
\caption{\label{fig:coolingrates}Cooling rate constraints from dark black hole detections combined with large scale structure collisions. Shaded regions correspond to the allowed parameter space in which (\textit{blue}) small halos at $z = 5, 10$ where DBHs form, (\textit{red}) energy loss fraction $f_{\rm lost} = 0.01, 0.1, 0.3$ (solid to dashed), for galaxy cluster collisions, and (\textit{brown}) $f_{\rm lost} = 0.5$ for dwarf galaxy collisions. We also include the cooling function (solid black line) for an example atomic dark matter model consistent with the interpretation that GW190425 is a binary DBH $(m_x = \SI{14}{\giga\electronvolt}, m_c = \SI{325}{\kilo\electronvolt}, \alpha_D = 0.01)$.
}
\end{figure}

To illustrate the constraint we chose $m_x = \SI{14}{\giga\electronvolt}$, consistent with the scenario that GW190425 is a binary DBH. In figure \ref{fig:coolingrates} we plot the above constraint on cooling rate $\Lambda$, across a range of temperatures corresponding to halo masses of $10^5 - \SI{e9}{\sol}$ at redshifts $z = 5$ and $z = 10$. For reference, free-fall times are $\mathcal{O}(\SI{0.1}{\giga\year})$ for these redshifts. \cite{PhysRevLett.120.051102}. The black line shows an atomic dark matter cooling curve for reference~\cite{PhysRevD.96.123001}.

To obtain an upper limit on cooling at high temperatures, we use observations of galaxy cluster collisions that indicate dark matter experiences minimal energy loss due to dark particle interactions~\cite{Markevitch:2003at}. For a cluster with dark matter surface density $\Sigma_{\rm s}$, colliding at velocity $v_{\rm coll}$ with another cluster, the cooling rate can be bounded by the maximum fraction of energy lost by the dark matter $f_{\rm lost}$, as,
\begin{align}
\begin{split}
	\Lambda \leq \left(\frac{f_{\rm lost} \mathcal{E}_{\rm k} v_{\rm coll}}{\Sigma_{\rm s}}\right)\left(\frac{m_x}{n_{\rm D}}\right)
\end{split}
\end{align}
Figure \ref{fig:coolingrates} shows how this constraint applies to cluster-scale temperatures if observational data bounds $f_{\rm lost} = 0.01, 0.1, 0.3$ (darkest to lightest). We use $v_{\rm coll}=\SI{1000}{\kilo\meter\per\second}$, $\Sigma_s = \SI{0.25}{\gram\per\centi\meter\squared}$ appropriate for the Bullet Cluster~\cite{Markevitch:2003at}. 

The lack of observed dwarf galaxies in contrast to the result from cosmological simulations - called ``the missing-satellite problem" - may suggest that dwarf galaxies are fragile under galaxy cluster collisions~\cite{Bullock:2017xww}. Interestingly, our analysis shows that some dwarf galaxies are likely to be disrupted by such collisions if dark matter is sufficiently dissipative to have formed the constituent DBHs of GW190425. Figure \ref{fig:coolingrates} shows a range of temperatures appropriate for dwarf galaxies with $\Sigma_{\rm s} =\SI{e-3}{\gram\per\centi\meter\squared}$, where collisions with $v_{\rm coll} = \SI{100}{\kilo\meter\per\second}$~\cite{Herrmann_2016,McConnachie_2012}, likely lead to disruption ($f_{\rm lost} \geq 0.5$). We show how the interactions that lead to cooling will also lead to the disruption of low temperature structures (dwarf galaxies) in collisions, without affecting large scale structure.

\section{Comparison with alternate dark matter compact object constraints}
Compact objects have long been considered as a possible component of the dark matter~\cite{Hawking1971, Chapline1975}, and by the 1990s interest in this candidate class peaked. Microlensing experiments that sought to measure signatures of compact objects in the galactic halo had promising early results that suggested a population of dark objects~\cite{Alcock:1993eu}. As LIGO and Virgo secured funding and broke ground, there was similar interest in the prospect of observing these objects via independent methods~\cite{Nakamura:1997sm}. Interest waned as microlensing surveys~\cite{Allsman:2000kg,Tisserand:2006zx}, gravitational-wave searches~\cite{Abbott:2005pf,PhysRevLett.121.231103,PhysRevLett.123.161102,Abbott:2007xi,wade2015}, and dwarf galaxy dynamics calculations~\cite{Brandt:2016aco,Koushiappas:2017chw} determined that compact objects are unlikely to make up all of the dark matter.

\begin{figure}
\includegraphics[width=\columnwidth]{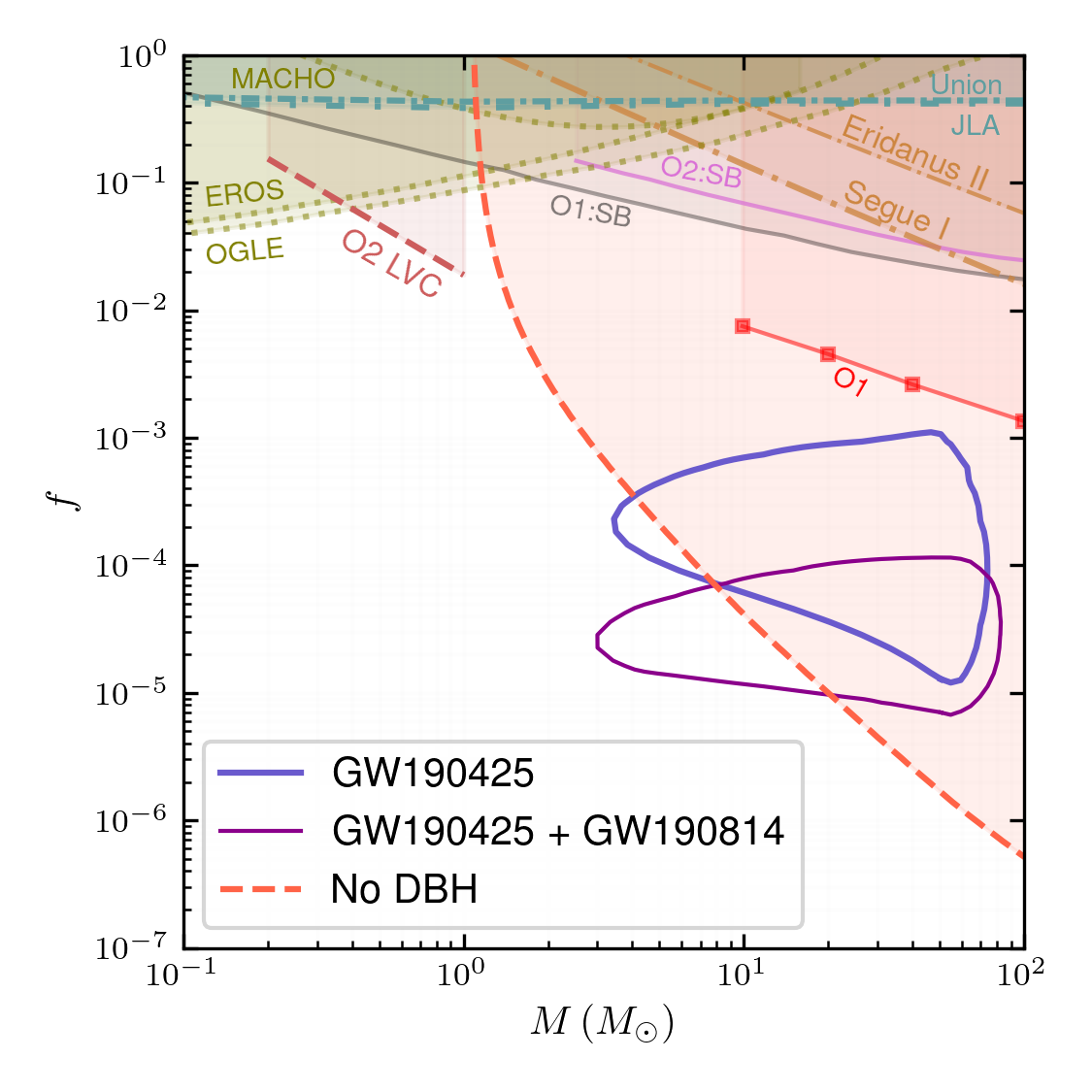}
\caption{\label{fig:CODM}Constraints on the fraction of dark matter in compact objects $f$ with our constraint on DBH overlaid. The DBH constraints are plotted for the mean of the mass distributions that extend over $M_{\textrm{min}} - M_{max}$. The O2 LVC constraints are specific to the PBH scenario of ~\cite{Nakamura:1997sm}. The microlensing constraints are derived from~\cite{Allsman:2000kg,Tisserand:2006zx,ogle}, and dwarf galaxy dynamics from~\cite{Brandt:2016aco,Koushiappas:2017chw}. The x-axis denotes the mass in solar masses of the dark matter object. PBH constraints are derived from~\cite{PhysRevLett.123.161102,Gaggero:2016dpq,Raidal:2017mfl,Raidal:2018bbj,Ali-Haimoud:2017rtz}. \textit{SB} stands for stochastic background.
}
\end{figure}

In the model we present, only a small fraction of the total dark matter cools and collapses to form DBH binaries. Indeed, the fraction of dark matter contained in DBHs is far below existing constraints on compact object dark matter. Figure \ref{fig:CODM} shows that if GW190425 and GW190814 are not DBH binaries, the resulting constraints on our model are $O(10^2 - 10^5)$ times stricter than limits from microlensing and dwarf galaxy dynamics. Conversely, if GW190425 is a DBH, then it exists in a region unconstrained by these observations. In figure \ref{fig:CODM}, we also show limits on $f_{\textrm{PBH}}$ from the non-observation of a stochastic background in the first and second observing runs of LIGO-Virgo reported in \cite{Raidal:2017mfl, Raidal:2018bbj}. \cite{Ali-Haimoud:2017rtz} reports potential upper limits on $f_{\rm PBH}$ for equal mass binaries with $M_{\rm PBH} \in [10,  300 \si{\sol}]$, also shown in figure \ref{fig:CODM}.
In interpreting figure \ref{fig:CODM}, note that the PBH constraints assume a delta function for initial black hole masses, except the O2 stochastic background constraint of \cite{Raidal:2018bbj} which assumes a lognormal mass function for PBHs, while in the DBH case the mass on the x-axis is the average mass over all initial mass distributions we considered for a given $M_{\textrm{min}}$ (see Fig. 1).  

\section{Conclusion}

Gravitational-wave astronomy can play a key role in understanding the properties of dissipative dark matter that can cool to form compact objects like DBHs. We illustrate how only a single gravitational-wave detection of a DBH binary opens up a novel framework for studying the properties of dissipative dark matter from gravitational wave astronomy. Interpreting GW190425 as a DBH binary not only constrains $f$ and $M_{\textrm{min}}$ for a DBH binary population predicted for atomic dark matter
in~\cite{Shandera:2018xkn}, but the constraints on $M_{\textrm{min}}$ also provide a direct upper limit on the heavy dark fermion mass $m_x$ through the Chandrasekhar limit (using $M_{\textrm DC}=M_{\textrm min}$. From the fact that the minimum of the dark-matter BH mass distribution is smaller than the usual Chandrasekhar limit of $1.4\,M_{\odot}$, we find that $m_x>0.95\,{\textrm GeV}$ with $99.9\%$ confidence. We also illustrate how an additional heavier DBH detection, GW190814, has little effect on the predicted $M_{\textrm{min}}$ but constrains $f$ to a higher extent. In this case, the Chandrasekhar limit allows for heavy dark fermion masses to be smaller than the proton mass.

Assuming a simple dark matter model forming bound states analogous to those of hydrogen, for which we adopt Population-III star formation and binary parameters for our DBH population, we also estimate atomic-physics parameters for dark matter. For example, if $M_{\textrm{DJ}} = \SI{0.92}{\sol}$ (the most probable value of $M_{\textrm min}$ if only GW190425 is a DBH), and assuming $m_x = \SI{14}{\giga\electronvolt}$, $m_c = \SI{325}{\kilo\electronvolt}$, and $\alpha_D = 0.01$, we estimate the molecular energy gap for these dark atoms to be \SI{2.2}{\milli\electronvolt}. If GW190425 is a DBH merger from a scenario similar to those we have studied here, additional small black holes would likely be observed by gravitational-wave detectors in the near future. In the future, the presence of an electromagnetic counterpart or the measurement of tidal deformability could each rule out the binary black hole scenario for events like GW190425. However, the observation of a population of sub-solar mass black holes would be clear evidence of a population of black holes that are distinct from standard formation mechanisms. This would need to be further studied and put into context of broader cosmological observables to determine the support for different scenarios such as the collapse of atomic dark matter that we discuss here, and others like primordial black holes, dark matter induced collapse of neutron stars, etc.

Assuming none of the LVC-detected events originate from dissipative dark matter allows a strong constraint on the fraction of dark matter in compact objects. We present the two scenarios of a DBH detection and the DBH null result mutually exclusively to illustrate concisely the information we can infer. This is a novel approach to constrain dark matter self-interactions, complimentary to studies of structure on the scale of galaxies and above. Especially if future gravitational-wave searches in the sub-solar mass regime do not find any compact object mergers, gravitational-wave data could be a means to rule out a significant range of dissipative dark matter models.

\appendix

\section{Computing posteriors}\label{appendix:posteriors}

A Bayesian approach is used to get the posterior probability of $f$ and $\overline{\theta} = \{M_{\textrm{min}}, r, b\}$. The probability is given by 
\begin{equation}
	P(f,\overline{\theta} | \overline{\mu}) 
	\propto
	P(f, \overline{\theta})
	P(\overline{\mu} | f, \overline{\theta}) 
\end{equation}
where $\overline{\mu} = \{\mu_i\}$ is a vector that corresponds to the mean number or expected value of DBH detections made by the LVC in pre-defined chirp mass bins denoted by $i$. The chirp masses range from $0.2-200 M_\odot$ and we define bins such that at most two LVC events lie in a bin. We choose these bins to correspond roughly to the parameter space searched by~\cite{Authors:2019qbw} and~\cite{Abbott:2020niy} and to simplify the marginalization over DBH versus astrophysical black hole mergers as described below. The expected value for event counts in a bin is given by $\mu_i = R_i(f,\overline{\theta}) \times (V_i(\overline{\theta})T)$, where $R_i(f, \overline{\theta})$ is the predicted rate of DBH mergers computed using equation~\ref{eq:rate}, and $V_i(\overline{\theta})T$ describes the sensitivity of the detectors to DBH mergers that fall in the given bin. Appendix~\ref{appendix:VT} outlines how we compute the $VT$ for this analysis.

To compute the likelihood, $\mathcal{L}(f,\overline{\theta} ; \overline{\mu}) \equiv P(\overline{\mu} | f, \overline{\theta})$, we first assume that the DBH event count in a particular chirp mass bin, $n_i$, follows a Poisson distribution such that $P(X=n_i) = \mu_i^{n_i}e^{-\mu_i}/n_i !$. We use this as the distribution for $\mu_i$ and compute the likelihood as
\begin{equation}\label{eq:likelihood}
	\mathcal{L}(f,\overline{\theta} ; \overline{\mu})
	= \displaystyle\prod_{i}
	 \frac{\int_{a}^{b}\mathcal{P}_i(\mu_i | f, \overline{\theta}) d\mu}
	 {\int_{0}^{\infty}\mathcal{P}_i(\mu_i | f, \overline{\theta}) d\mu}
\end{equation}
where we integrate from the tail of the distribution to $\mu_i$. When $n_i$(DBH) =1, $(a, b) = (0, \infty)$ if $\mu_i < 1$, and $(a,b) = (\mu_i, \infty)$ if $\mu_i > 1$. For all other cases, we integrate over $(a,b) = (\mu_i, \infty)$.

For the other LVC events, $\mathcal{P}(\mu_i | f, \overline{\theta})$ includes marginalization over the probability that the event could be a DBH or an astrophysical binary black-hole merger, reflecting that the LVC events could be from either kind of compact object with equal probability. The $\mu_i$ distribution is given by
\begin{equation}
	\mathcal{P}(\mu_i)
	=
	\left(\frac{1+\mu_i}{2}\right)e^{\mu_i}
\end{equation}
for a single event in the bin, and
\begin{equation}
	\mathcal{P}(\mu_i)
	=
	\left(\frac{2+4\mu_i+\mu_i^2}{8}\right)
	e^{\mu_i}
\end{equation}
for two events in the bin. The likelihood is computed using equation~\ref{eq:likelihood} with $(a,b) = (\mu_i, \infty)$. 
\begin{figure}[ht]
  \includegraphics[width=\columnwidth]{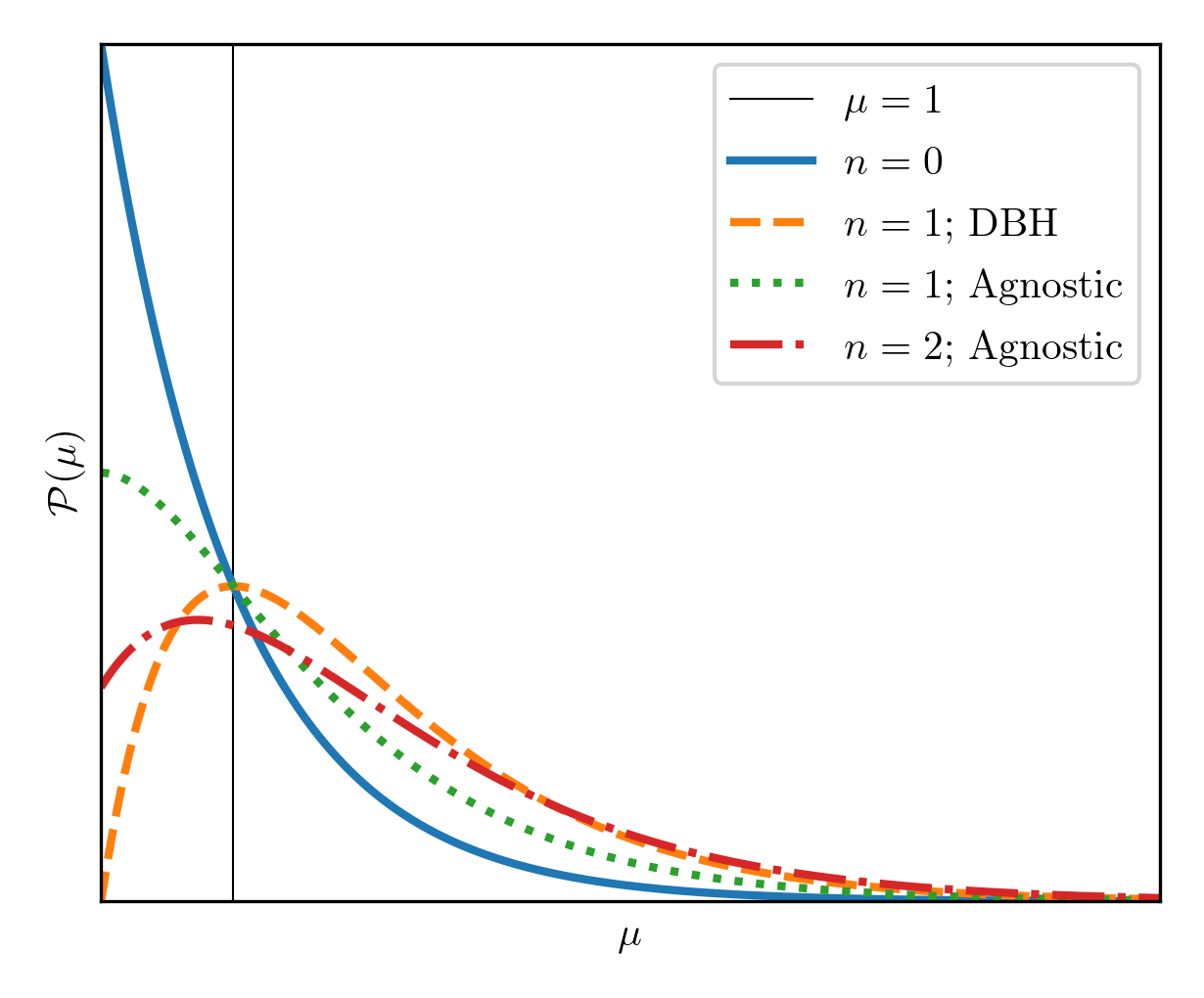}
  \caption{Distributions for the expected number of DBH events, $\mu$ for varying number of observed LVC binary black hole coalescences, $n$. The agnostic cases indicate that the observed events could be either astrophysical or dark binary black hole coalescences with equal probability.}
\label{fig:ratepos}
\end{figure}

\begin{figure*}
  \includegraphics[width=\textwidth]{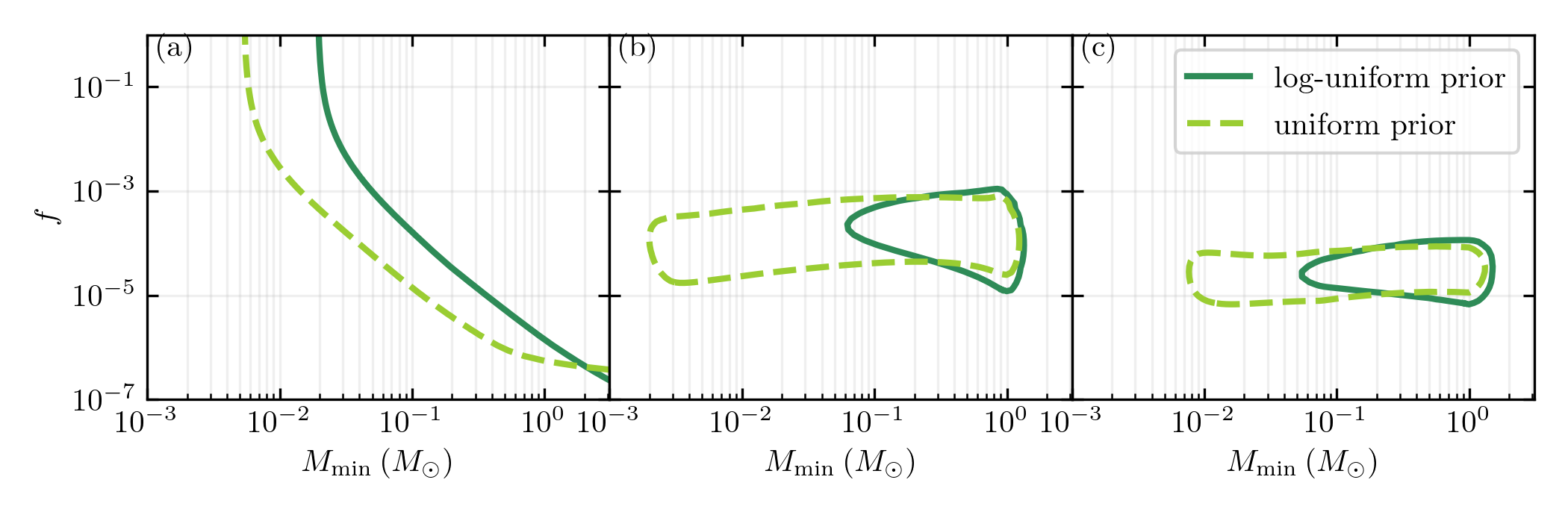}
  \caption{Constraints on $f$ and $M_{\textrm{min}}$ using a log-uniform prior(\textit{solid}) and uniform prior(\textit{dashed}) on $r=M_{\textrm{max}}/M_{\textrm{min}}$ for the three cases reported in figure~\ref{fig:constraints} - (a) no observed DBH events, (b) GW190425 is a DBH event, and (c) GW190425 and GW190814 are DBH events.}
\label{fig:priors}
\end{figure*}

We choose uninformative priors for $\{f, \overline{\theta}\}$ in the analysis. We assume a uniform prior for $M_{\textrm{min}} \in [10^{-3},  3.1]$, $f \in [10^{-10}, 1]$ and $b \in [-1, 2]$, and a log-uniform prior for $r \in [2, 10^3]$. The log-uniform prior equitably weights $r$ unlike a uniform prior. Figure~\ref{fig:priors} shows that the constraint on $M_{\textrm{min}}$ extends to lower masses if we choose a uniform prior on $r$, exhibiting a preference for high values of $r$. However, there is no evidence for such preference. The constraints on $M_{\textrm{min}}$ are also observed to be more sensitive to the lower bound on $r$. Therefore, we report constraints using a log-uniform prior for $r$ in section~\ref{sec:paramConstraints}, with the most conservative bounds on the prior.

\section{Sensitive Volume}\label{appendix:VT}
The sensitive volume, $VT$, is computed over the range of chirp masses $0.2 - 200\ M_\odot$ in each chirp mass bin denoted by $i$ for mass-ratios, $q\in[0,1]$ using 
\begin{equation}\label{eq:VT1}
	V_i T (q, \mathcal{M} = m_i) = (0.088)\frac{4\pi}{3}\sum^3_{j=1}D^3_j(q, m_i)T_j	
\end{equation}
where $D_j(q,m_i)$ is the horizon distance computed for an SNR threshold of 8 in the LIGO Livingston detector during the observing run $j =1,2,3$ that spanned over observation times $T_j$ respectively. 

\begin{figure*}
  \includegraphics[width=\textwidth]{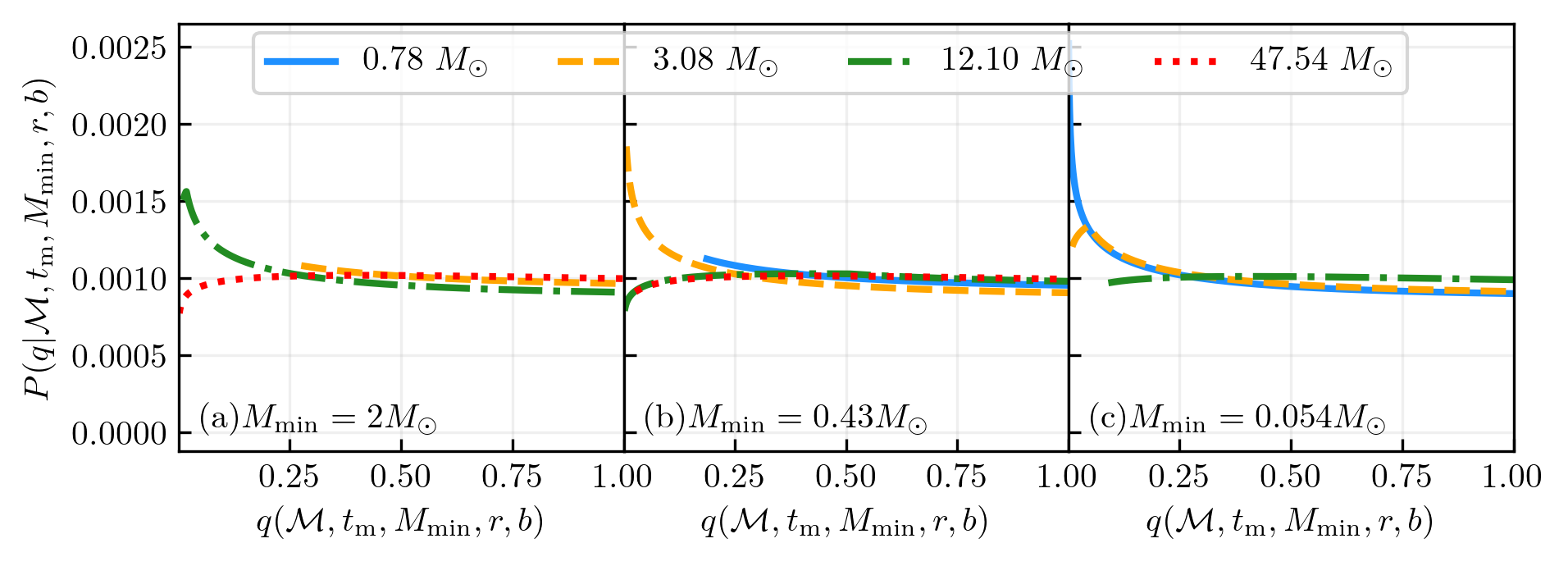}
  \caption{\label{fig:q_dist}Unique distributions of mass ratio, $q$ given some chirp mass for different values of $M_{\textrm{min}}$. The slope of the initial mass distribution, $b = 2$ and ratio, $r = 1000$ for each subfigure. The distributions show binaries that merge within 10 Gyr for a given chirp mass.}
\end{figure*}

We compute a weighted average of the $VT$s over all possible binaries for a given chirp mass within mass ranges representative of the search parameter space used for compact binary coalescences conducted in data from the second observing run of LVC. The parameter space places hard cuts on the binary systems that contribute to the $VT$ integral,
\begin{equation}\label{eq:VT2}
	V_iT(\mathcal{M}=m_i | \overline{\theta}) = \int_{q_{\textrm{min}}}^1 \mathcal{P}(q|m_i,  t_{\textrm{m}}, \overline{\theta})V_i T(q,m_i|\overline{\theta})dq
\end{equation}
Here, $\mathcal{P}(q | m_i, t_{\textrm{m}}, \overline{\theta})$ is the distribution of mass-ratios, $q \in [q_{\textrm{min}}, 1]$  for a given chirp mass dictated by the choice of parameters, $\overline{\theta}$ and $t_{\textrm{m}} = 10$ Gyr. This is different from the initial distribution of mass-ratios which is independent of model parameters. Figure~\ref{fig:q_dist} shows how the allowed values for $q$ change with chirp mass and $\overline{\theta}$. For a given $M_{\textrm{min}}$, possible DBH chirp masses range from $2^{-1/5}M_{\textrm{min}}$ to $2^{-1/5}rM_{\textrm{min}}$. This is evident in figure~\ref{fig:q_dist} where no binaries occur for $\mathcal{M} = 0.78 M_\odot $ when $M_{\textrm{min}} = 2 M_\odot$, as well as for  $\mathcal{M} = 47.54 M_\odot $ when $M_{\textrm{min}} = 0.054 M_\odot$ because the chirp masses lie outside the permissible range for the respective values of $M_{\textrm{min}}$.

The left panel of figure~\ref{fig:VT_R} illustrates how these weighted $VT$s vary over chirp mass for various values of $M_{\textrm{min}}$ for some fixed values of $b$ ad $r$.

\begin{acknowledgments}
We thank Sarah Caudill for reviewing this manuscript for the LIGO/Virgo collaborations and the LIGO/Virgo Rates and Populations working group for helpful feedback. This document has been assigned the LIGO document number LIGO-P2000291. Funding for this work was provided by the Charles E. Kaufman Foundation of the Pittsburgh Foundation. T. A. was supported through the Eberly College of Science Office of Science Engagement. Computing resources and personnel were provided by the Pennsylvania State University. This work was supported by the National Science Foundation through PHY-2011865.  The authors are grateful for computational resources provided by the LIGO Laboratory and supported by National Science Foundation Grants PHY-0757058 and PHY-0823459. 
\end{acknowledgments}


\bibliography{references}

\begin{thebibliography}{73}%
\makeatletter
\providecommand \@ifxundefined [1]{%
 \@ifx{#1\undefined}
}%
\providecommand \@ifnum [1]{%
 \ifnum #1\expandafter \@firstoftwo
 \else \expandafter \@secondoftwo
 \fi
}%
\providecommand \@ifx [1]{%
 \ifx #1\expandafter \@firstoftwo
 \else \expandafter \@secondoftwo
 \fi
}%
\providecommand \natexlab [1]{#1}%
\providecommand \enquote  [1]{``#1''}%
\providecommand \bibnamefont  [1]{#1}%
\providecommand \bibfnamefont [1]{#1}%
\providecommand \citenamefont [1]{#1}%
\providecommand \href@noop [0]{\@secondoftwo}%
\providecommand \href [0]{\begingroup \@sanitize@url \@href}%
\providecommand \@href[1]{\@@startlink{#1}\@@href}%
\providecommand \@@href[1]{\endgroup#1\@@endlink}%
\providecommand \@sanitize@url [0]{\catcode `\\12\catcode `\$12\catcode
  `\&12\catcode `\#12\catcode `\^12\catcode `\_12\catcode `\%12\relax}%
\providecommand \@@startlink[1]{}%
\providecommand \@@endlink[0]{}%
\providecommand \url  [0]{\begingroup\@sanitize@url \@url }%
\providecommand \@url [1]{\endgroup\@href {#1}{\urlprefix }}%
\providecommand \urlprefix  [0]{URL }%
\providecommand \Eprint [0]{\href }%
\providecommand \doibase [0]{http://dx.doi.org/}%
\providecommand \selectlanguage [0]{\@gobble}%
\providecommand \bibinfo  [0]{\@secondoftwo}%
\providecommand \bibfield  [0]{\@secondoftwo}%
\providecommand \translation [1]{[#1]}%
\providecommand \BibitemOpen [0]{}%
\providecommand \bibitemStop [0]{}%
\providecommand \bibitemNoStop [0]{.\EOS\space}%
\providecommand \EOS [0]{\spacefactor3000\relax}%
\providecommand \BibitemShut  [1]{\csname bibitem#1\endcsname}%
\let\auto@bib@innerbib\@empty
\bibitem [{\citenamefont {Hoyle}(1954)}]{Hoyle:1954zz}%
  \BibitemOpen
  \bibfield  {author} {\bibinfo {author} {\bibfnamefont {F.}~\bibnamefont
  {Hoyle}},\ }\href {\doibase 10.1086/190005} {\bibfield  {journal} {\bibinfo
  {journal} {Astrophys. J. Suppl.}\ }\textbf {\bibinfo {volume} {1}},\ \bibinfo
  {pages} {121} (\bibinfo {year} {1954})}\BibitemShut {NoStop}%
\bibitem [{\citenamefont {{Planck Collaboration:}}\ \emph
  {et~al.}(2015)\citenamefont {{Planck Collaboration:}}, \citenamefont {{Ade,
  P. A. R.}},\ and\ \citenamefont {{others}}}]{refId0}%
  \BibitemOpen
  \bibfield  {author} {\bibinfo {author} {\bibnamefont {{Planck
  Collaboration:}}}, \bibinfo {author} {\bibnamefont {{Ade, P. A. R.}}}, \ and\
  \bibinfo {author} {\bibnamefont {{others}}},\ }\href {\doibase
  10.1051/0004-6361/201424496} {\bibfield  {journal} {\bibinfo  {journal}
  {A\&A}\ }\textbf {\bibinfo {volume} {580}},\ \bibinfo {pages} {A22} (\bibinfo
  {year} {2015})}\BibitemShut {NoStop}%
\bibitem [{\citenamefont {Arnett}\ \emph {et~al.}(1989)\citenamefont {Arnett},
  \citenamefont {Bahcall}, \citenamefont {Kirshner},\ and\ \citenamefont
  {Woosley}}]{doi:10.1146/annurev.aa.27.090189.003213}%
  \BibitemOpen
  \bibfield  {author} {\bibinfo {author} {\bibfnamefont {W.~D.}\ \bibnamefont
  {Arnett}}, \bibinfo {author} {\bibfnamefont {J.~N.}\ \bibnamefont {Bahcall}},
  \bibinfo {author} {\bibfnamefont {R.~P.}\ \bibnamefont {Kirshner}}, \ and\
  \bibinfo {author} {\bibfnamefont {S.~E.}\ \bibnamefont {Woosley}},\ }\href
  {\doibase 10.1146/annurev.aa.27.090189.003213} {\bibfield  {journal}
  {\bibinfo  {journal} {Annual Review of Astronomy and Astrophysics}\ }\textbf
  {\bibinfo {volume} {27}},\ \bibinfo {pages} {629} (\bibinfo {year} {1989})},\
  \Eprint
  {http://arxiv.org/abs/https://doi.org/10.1146/annurev.aa.27.090189.003213}
  {https://doi.org/10.1146/annurev.aa.27.090189.003213} \BibitemShut {NoStop}%
\bibitem [{\citenamefont {Penning}(2018)}]{Penning_2018}%
  \BibitemOpen
  \bibfield  {author} {\bibinfo {author} {\bibfnamefont {B.}~\bibnamefont
  {Penning}},\ }\href {\doibase 10.1088/1361-6471/aabea7} {\bibfield  {journal}
  {\bibinfo  {journal} {Journal of Physics G: Nuclear and Particle Physics}\
  }\textbf {\bibinfo {volume} {45}},\ \bibinfo {pages} {063001} (\bibinfo
  {year} {2018})}\BibitemShut {NoStop}%
\bibitem [{\citenamefont {Dainese}\ \emph {et~al.}(2019)\citenamefont
  {Dainese}, \citenamefont {Mangano}, \citenamefont {Meyer}, \citenamefont
  {Nisati}, \citenamefont {Salam},\ and\ \citenamefont
  {Vesterinen}}]{Dainese:2019rgk}%
  \BibitemOpen
  \bibinfo {editor} {\bibfnamefont {A.}~\bibnamefont {Dainese}}, \bibinfo
  {editor} {\bibfnamefont {M.}~\bibnamefont {Mangano}}, \bibinfo {editor}
  {\bibfnamefont {A.~B.}\ \bibnamefont {Meyer}}, \bibinfo {editor}
  {\bibfnamefont {A.}~\bibnamefont {Nisati}}, \bibinfo {editor} {\bibfnamefont
  {G.}~\bibnamefont {Salam}}, \ and\ \bibinfo {editor} {\bibfnamefont {M.~A.}\
  \bibnamefont {Vesterinen}},\ eds.,\ \href {\doibase 10.23731/CYRM-2019-007}
  {\emph {\bibinfo {title} {{Report on the Physics at the HL-LHC,and
  Perspectives for the HE-LHC}}}},\ \bibinfo {series} {CERN Yellow Reports:
  Monographs}, Vol.\ \bibinfo {volume} {7/2019}\ (\bibinfo  {publisher}
  {CERN},\ \bibinfo {address} {Geneva, Switzerland},\ \bibinfo {year}
  {2019})\BibitemShut {NoStop}%
\bibitem [{\citenamefont {Akerib}\ \emph {et~al.}(2020)\citenamefont {Akerib}
  \emph {et~al.}}]{PhysRevD.101.052002}%
  \BibitemOpen
  \bibfield  {author} {\bibinfo {author} {\bibfnamefont {D.~S.}\ \bibnamefont
  {Akerib}} \emph {et~al.} (\bibinfo {collaboration} {LUX-ZEPLIN
  Collaboration}),\ }\href {\doibase 10.1103/PhysRevD.101.052002} {\bibfield
  {journal} {\bibinfo  {journal} {Phys. Rev. D}\ }\textbf {\bibinfo {volume}
  {101}},\ \bibinfo {pages} {052002} (\bibinfo {year} {2020})}\BibitemShut
  {NoStop}%
\bibitem [{\citenamefont {Aguilar-Arevalo}\ \emph {et~al.}(2017)\citenamefont
  {Aguilar-Arevalo} \emph {et~al.}}]{PhysRevLett.118.221803}%
  \BibitemOpen
  \bibfield  {author} {\bibinfo {author} {\bibnamefont {Aguilar-Arevalo}} \emph
  {et~al.} (\bibinfo {collaboration} {MiniBooNE-DM Collaboration}),\ }\href
  {\doibase 10.1103/PhysRevLett.118.221803} {\bibfield  {journal} {\bibinfo
  {journal} {Phys. Rev. Lett.}\ }\textbf {\bibinfo {volume} {118}},\ \bibinfo
  {pages} {221803} (\bibinfo {year} {2017})}\BibitemShut {NoStop}%
\bibitem [{\citenamefont {Diamond}(2019)}]{Diamond:2019luc}%
  \BibitemOpen
  \bibfield  {author} {\bibinfo {author} {\bibfnamefont {M.}~\bibnamefont
  {Diamond}} (\bibinfo {collaboration} {MATHUSLA}),\ }\href {\doibase
  10.22323/1.367.0115} {\bibfield  {journal} {\bibinfo  {journal} {PoS}\
  }\textbf {\bibinfo {volume} {LeptonPhoton2019}},\ \bibinfo {pages} {115}
  (\bibinfo {year} {2019})}\BibitemShut {NoStop}%
\bibitem [{\citenamefont {Roszkowski}\ \emph {et~al.}(2018)\citenamefont
  {Roszkowski}, \citenamefont {Sessolo},\ and\ \citenamefont
  {Trojanowski}}]{Roszkowski_2018}%
  \BibitemOpen
  \bibfield  {author} {\bibinfo {author} {\bibfnamefont {L.}~\bibnamefont
  {Roszkowski}}, \bibinfo {author} {\bibfnamefont {E.~M.}\ \bibnamefont
  {Sessolo}}, \ and\ \bibinfo {author} {\bibfnamefont {S.}~\bibnamefont
  {Trojanowski}},\ }\href {\doibase 10.1088/1361-6633/aab913} {\bibfield
  {journal} {\bibinfo  {journal} {Reports on Progress in Physics}\ }\textbf
  {\bibinfo {volume} {81}},\ \bibinfo {pages} {066201} (\bibinfo {year}
  {2018})}\BibitemShut {NoStop}%
\bibitem [{\citenamefont {Conrad}\ and\ \citenamefont
  {Reimer}(2017)}]{Conrad:2017pms}%
  \BibitemOpen
  \bibfield  {author} {\bibinfo {author} {\bibfnamefont {J.}~\bibnamefont
  {Conrad}}\ and\ \bibinfo {author} {\bibfnamefont {O.}~\bibnamefont
  {Reimer}},\ }\href {\doibase 10.1038/nphys4049} {\bibfield  {journal}
  {\bibinfo  {journal} {Nature Phys.}\ }\textbf {\bibinfo {volume} {13}},\
  \bibinfo {pages} {224} (\bibinfo {year} {2017})},\ \Eprint
  {http://arxiv.org/abs/1705.11165} {arXiv:1705.11165 [astro-ph.HE]}
  \BibitemShut {NoStop}%
\bibitem [{\citenamefont {Aasi}\ \emph {et~al.}(2015)\citenamefont {Aasi} \emph
  {et~al.}}]{TheLIGOScientific:2014jea}%
  \BibitemOpen
  \bibfield  {author} {\bibinfo {author} {\bibfnamefont {J.}~\bibnamefont
  {Aasi}} \emph {et~al.} (\bibinfo {collaboration} {LIGO Scientific}),\ }\href
  {\doibase 10.1088/0264-9381/32/7/074001} {\bibfield  {journal} {\bibinfo
  {journal} {Class. Quant. Grav.}\ }\textbf {\bibinfo {volume} {32}},\ \bibinfo
  {pages} {074001} (\bibinfo {year} {2015})},\ \Eprint
  {http://arxiv.org/abs/1411.4547} {arXiv:1411.4547 [gr-qc]} \BibitemShut
  {NoStop}%
\bibitem [{\citenamefont {Acernese}\ \emph {et~al.}(2015)\citenamefont
  {Acernese} \emph {et~al.}}]{TheVirgo:2014hva}%
  \BibitemOpen
  \bibfield  {author} {\bibinfo {author} {\bibfnamefont {F.}~\bibnamefont
  {Acernese}} \emph {et~al.} (\bibinfo {collaboration} {VIRGO}),\ }\href
  {\doibase 10.1088/0264-9381/32/2/024001} {\bibfield  {journal} {\bibinfo
  {journal} {Class. Quant. Grav.}\ }\textbf {\bibinfo {volume} {32}},\ \bibinfo
  {pages} {024001} (\bibinfo {year} {2015})},\ \Eprint
  {http://arxiv.org/abs/1408.3978} {arXiv:1408.3978 [gr-qc]} \BibitemShut
  {NoStop}%
\bibitem [{\citenamefont {Abbott}\ \emph {et~al.}(2016)\citenamefont {Abbott}
  \emph {et~al.}}]{Abbott:2016blz}%
  \BibitemOpen
  \bibfield  {author} {\bibinfo {author} {\bibfnamefont {B.~P.}\ \bibnamefont
  {Abbott}} \emph {et~al.} (\bibinfo {collaboration} {Virgo, LIGO
  Scientific}),\ }\href {\doibase 10.1103/PhysRevLett.116.061102} {\bibfield
  {journal} {\bibinfo  {journal} {Phys. Rev. Lett.}\ }\textbf {\bibinfo
  {volume} {116}},\ \bibinfo {pages} {061102} (\bibinfo {year} {2016})},\
  \Eprint {http://arxiv.org/abs/1602.03837} {arXiv:1602.03837 [gr-qc]}
  \BibitemShut {NoStop}%
\bibitem [{\citenamefont {Carr}\ \emph {et~al.}(2016)\citenamefont {Carr},
  \citenamefont {K{\"{u}}hnel},\ and\ \citenamefont {Sandstad}}]{Carr2016}%
  \BibitemOpen
  \bibfield  {author} {\bibinfo {author} {\bibfnamefont {B.}~\bibnamefont
  {Carr}}, \bibinfo {author} {\bibfnamefont {F.}~\bibnamefont {K{\"{u}}hnel}},
  \ and\ \bibinfo {author} {\bibfnamefont {M.}~\bibnamefont {Sandstad}},\
  }\href {\doibase 10.1103/PhysRevD.94.083504} {\bibfield  {journal} {\bibinfo
  {journal} {Physical Review D}\ }\textbf {\bibinfo {volume} {94}},\ \bibinfo
  {pages} {083504} (\bibinfo {year} {2016})},\ \Eprint
  {http://arxiv.org/abs/1607.06077} {arXiv:1607.06077} \BibitemShut {NoStop}%
\bibitem [{\citenamefont {Bird}\ \emph {et~al.}(2016)\citenamefont {Bird},
  \citenamefont {Cholis}, \citenamefont {Mu{\~n}oz}, \citenamefont
  {Ali-Ha{\"\i}moud}, \citenamefont {Kamionkowski}, \citenamefont {Kovetz},
  \citenamefont {Raccanelli},\ and\ \citenamefont {Riess}}]{Bird:2016dcv}%
  \BibitemOpen
  \bibfield  {author} {\bibinfo {author} {\bibfnamefont {S.}~\bibnamefont
  {Bird}}, \bibinfo {author} {\bibfnamefont {I.}~\bibnamefont {Cholis}},
  \bibinfo {author} {\bibfnamefont {J.~B.}\ \bibnamefont {Mu{\~n}oz}}, \bibinfo
  {author} {\bibfnamefont {Y.}~\bibnamefont {Ali-Ha{\"\i}moud}}, \bibinfo
  {author} {\bibfnamefont {M.}~\bibnamefont {Kamionkowski}}, \bibinfo {author}
  {\bibfnamefont {E.~D.}\ \bibnamefont {Kovetz}}, \bibinfo {author}
  {\bibfnamefont {A.}~\bibnamefont {Raccanelli}}, \ and\ \bibinfo {author}
  {\bibfnamefont {A.~G.}\ \bibnamefont {Riess}},\ }\href {\doibase
  10.1103/PhysRevLett.116.201301} {\bibfield  {journal} {\bibinfo  {journal}
  {Phys. Rev. Lett.}\ }\textbf {\bibinfo {volume} {116}},\ \bibinfo {pages}
  {201301} (\bibinfo {year} {2016})},\ \Eprint
  {http://arxiv.org/abs/1603.00464} {arXiv:1603.00464 [astro-ph.CO]}
  \BibitemShut {NoStop}%
\bibitem [{\citenamefont {Sasaki}\ \emph {et~al.}(2016)\citenamefont {Sasaki},
  \citenamefont {Suyama}, \citenamefont {Tanaka},\ and\ \citenamefont
  {Yokoyama}}]{Sasaki:2016jop}%
  \BibitemOpen
  \bibfield  {author} {\bibinfo {author} {\bibfnamefont {M.}~\bibnamefont
  {Sasaki}}, \bibinfo {author} {\bibfnamefont {T.}~\bibnamefont {Suyama}},
  \bibinfo {author} {\bibfnamefont {T.}~\bibnamefont {Tanaka}}, \ and\ \bibinfo
  {author} {\bibfnamefont {S.}~\bibnamefont {Yokoyama}},\ }\href {\doibase
  10.1103/PhysRevLett.117.061101} {\bibfield  {journal} {\bibinfo  {journal}
  {Phys. Rev. Lett.}\ }\textbf {\bibinfo {volume} {117}},\ \bibinfo {pages}
  {061101} (\bibinfo {year} {2016})},\ \Eprint
  {http://arxiv.org/abs/1603.08338} {arXiv:1603.08338 [astro-ph.CO]}
  \BibitemShut {NoStop}%
\bibitem [{\citenamefont {Clesse}\ and\ \citenamefont
  {Garcia-Bellido}(2020)}]{Clesse:2020ghq}%
  \BibitemOpen
  \bibfield  {author} {\bibinfo {author} {\bibfnamefont {S.}~\bibnamefont
  {Clesse}}\ and\ \bibinfo {author} {\bibfnamefont {J.}~\bibnamefont
  {Garcia-Bellido}},\ }\href@noop {} {\  (\bibinfo {year} {2020})},\ \Eprint
  {http://arxiv.org/abs/2007.06481} {arXiv:2007.06481 [astro-ph.CO]}
  \BibitemShut {NoStop}%
\bibitem [{\citenamefont {Abbott}\ \emph {et~al.}(2005)\citenamefont {Abbott}
  \emph {et~al.}}]{Abbott:2005pf}%
  \BibitemOpen
  \bibfield  {author} {\bibinfo {author} {\bibfnamefont {B.}~\bibnamefont
  {Abbott}} \emph {et~al.} (\bibinfo {collaboration} {LIGO Scientific}),\
  }\href {\doibase 10.1103/PhysRevD.72.082002} {\bibfield  {journal} {\bibinfo
  {journal} {Phys. Rev.}\ }\textbf {\bibinfo {volume} {D72}},\ \bibinfo {pages}
  {082002} (\bibinfo {year} {2005})},\ \Eprint
  {http://arxiv.org/abs/gr-qc/0505042} {arXiv:gr-qc/0505042 [gr-qc]}
  \BibitemShut {NoStop}%
\bibitem [{\citenamefont {Abbott}\ \emph {et~al.}(2018)\citenamefont {Abbott}
  \emph {et~al.}}]{PhysRevLett.121.231103}%
  \BibitemOpen
  \bibfield  {author} {\bibinfo {author} {\bibfnamefont {B.~P.}\ \bibnamefont
  {Abbott}} \emph {et~al.} (\bibinfo {collaboration} {LIGO Scientific
  Collaboration and Virgo Collaboration}),\ }\href {\doibase
  10.1103/PhysRevLett.121.231103} {\bibfield  {journal} {\bibinfo  {journal}
  {Phys. Rev. Lett.}\ }\textbf {\bibinfo {volume} {121}},\ \bibinfo {pages}
  {231103} (\bibinfo {year} {2018})}\BibitemShut {NoStop}%
\bibitem [{\citenamefont {Abbott}\ \emph
  {et~al.}(2019{\natexlab{a}})\citenamefont {Abbott} \emph
  {et~al.}}]{PhysRevLett.123.161102}%
  \BibitemOpen
  \bibfield  {author} {\bibinfo {author} {\bibfnamefont {B.~P.}\ \bibnamefont
  {Abbott}} \emph {et~al.} (\bibinfo {collaboration} {LIGO Scientific
  Collaboration and the Virgo Collaboration}),\ }\href {\doibase
  10.1103/PhysRevLett.123.161102} {\bibfield  {journal} {\bibinfo  {journal}
  {Phys. Rev. Lett.}\ }\textbf {\bibinfo {volume} {123}},\ \bibinfo {pages}
  {161102} (\bibinfo {year} {2019}{\natexlab{a}})}\BibitemShut {NoStop}%
\bibitem [{\citenamefont {Ebersold}\ and\ \citenamefont
  {Tiwari}(2020)}]{PhysRevD.101.104041}%
  \BibitemOpen
  \bibfield  {author} {\bibinfo {author} {\bibfnamefont {M.}~\bibnamefont
  {Ebersold}}\ and\ \bibinfo {author} {\bibfnamefont {S.}~\bibnamefont
  {Tiwari}},\ }\href {\doibase 10.1103/PhysRevD.101.104041} {\bibfield
  {journal} {\bibinfo  {journal} {Phys. Rev. D}\ }\textbf {\bibinfo {volume}
  {101}},\ \bibinfo {pages} {104041} (\bibinfo {year} {2020})}\BibitemShut
  {NoStop}%
\bibitem [{\citenamefont {Bertone}\ and\ \citenamefont
  {Tait}(2018)}]{Bertone:2018krk}%
  \BibitemOpen
  \bibfield  {author} {\bibinfo {author} {\bibfnamefont {G.}~\bibnamefont
  {Bertone}}\ and\ \bibinfo {author} {\bibfnamefont {M.}~\bibnamefont {Tait},
  \bibfnamefont {Tim}},\ }\href {\doibase 10.1038/s41586-018-0542-z} {\bibfield
   {journal} {\bibinfo  {journal} {Nature}\ }\textbf {\bibinfo {volume}
  {562}},\ \bibinfo {pages} {51} (\bibinfo {year} {2018})},\ \Eprint
  {http://arxiv.org/abs/1810.01668} {arXiv:1810.01668 [astro-ph.CO]}
  \BibitemShut {NoStop}%
\bibitem [{\citenamefont {Bullock}\ and\ \citenamefont
  {Boylan-Kolchin}(2017)}]{Bullock:2017xww}%
  \BibitemOpen
  \bibfield  {author} {\bibinfo {author} {\bibfnamefont {J.~S.}\ \bibnamefont
  {Bullock}}\ and\ \bibinfo {author} {\bibfnamefont {M.}~\bibnamefont
  {Boylan-Kolchin}},\ }\href {\doibase 10.1146/annurev-astro-091916-055313}
  {\bibfield  {journal} {\bibinfo  {journal} {Ann. Rev. Astron. Astrophys.}\
  }\textbf {\bibinfo {volume} {55}},\ \bibinfo {pages} {343} (\bibinfo {year}
  {2017})},\ \Eprint {http://arxiv.org/abs/1707.04256} {arXiv:1707.04256
  [astro-ph.CO]} \BibitemShut {NoStop}%
\bibitem [{\citenamefont {Essig}\ \emph {et~al.}(2019)\citenamefont {Essig},
  \citenamefont {Mcdermott}, \citenamefont {Yu},\ and\ \citenamefont
  {Zhong}}]{Essig:2018pzq}%
  \BibitemOpen
  \bibfield  {author} {\bibinfo {author} {\bibfnamefont {R.}~\bibnamefont
  {Essig}}, \bibinfo {author} {\bibfnamefont {S.~D.}\ \bibnamefont
  {Mcdermott}}, \bibinfo {author} {\bibfnamefont {H.-B.}\ \bibnamefont {Yu}}, \
  and\ \bibinfo {author} {\bibfnamefont {Y.-M.}\ \bibnamefont {Zhong}},\ }\href
  {\doibase 10.1103/PhysRevLett.123.121102} {\bibfield  {journal} {\bibinfo
  {journal} {Phys. Rev. Lett.}\ }\textbf {\bibinfo {volume} {123}},\ \bibinfo
  {pages} {121102} (\bibinfo {year} {2019})},\ \Eprint
  {http://arxiv.org/abs/1809.01144} {arXiv:1809.01144 [hep-ph]} \BibitemShut
  {NoStop}%
\bibitem [{\citenamefont {Choquette}\ \emph {et~al.}(2019)\citenamefont
  {Choquette}, \citenamefont {Cline},\ and\ \citenamefont
  {Cornell}}]{Choquette:2018lvq}%
  \BibitemOpen
  \bibfield  {author} {\bibinfo {author} {\bibfnamefont {J.}~\bibnamefont
  {Choquette}}, \bibinfo {author} {\bibfnamefont {J.~M.}\ \bibnamefont
  {Cline}}, \ and\ \bibinfo {author} {\bibfnamefont {J.~M.}\ \bibnamefont
  {Cornell}},\ }\href {\doibase 10.1088/1475-7516/2019/07/036} {\bibfield
  {journal} {\bibinfo  {journal} {JCAP}\ }\textbf {\bibinfo {volume} {07}},\
  \bibinfo {pages} {036} (\bibinfo {year} {2019})},\ \Eprint
  {http://arxiv.org/abs/1812.05088} {arXiv:1812.05088 [astro-ph.CO]}
  \BibitemShut {NoStop}%
\bibitem [{\citenamefont {Latif}\ \emph {et~al.}(2019)\citenamefont {Latif},
  \citenamefont {Lupi}, \citenamefont {Schleicher}, \citenamefont {D'Amico},
  \citenamefont {Panci},\ and\ \citenamefont {Bovino}}]{Latif:2018kqv}%
  \BibitemOpen
  \bibfield  {author} {\bibinfo {author} {\bibfnamefont {M.}~\bibnamefont
  {Latif}}, \bibinfo {author} {\bibfnamefont {A.}~\bibnamefont {Lupi}},
  \bibinfo {author} {\bibfnamefont {D.}~\bibnamefont {Schleicher}}, \bibinfo
  {author} {\bibfnamefont {G.}~\bibnamefont {D'Amico}}, \bibinfo {author}
  {\bibfnamefont {P.}~\bibnamefont {Panci}}, \ and\ \bibinfo {author}
  {\bibfnamefont {S.}~\bibnamefont {Bovino}},\ }\href {\doibase
  10.1093/mnras/stz608} {\bibfield  {journal} {\bibinfo  {journal} {Mon. Not.
  Roy. Astron. Soc.}\ }\textbf {\bibinfo {volume} {485}},\ \bibinfo {pages}
  {3352} (\bibinfo {year} {2019})},\ \Eprint {http://arxiv.org/abs/1812.03104}
  {arXiv:1812.03104 [astro-ph.CO]} \BibitemShut {NoStop}%
\bibitem [{\citenamefont {D'Amico}\ \emph {et~al.}(2018)\citenamefont
  {D'Amico}, \citenamefont {Panci}, \citenamefont {Lupi}, \citenamefont
  {Bovino},\ and\ \citenamefont {Silk}}]{DAmico:2017lqj}%
  \BibitemOpen
  \bibfield  {author} {\bibinfo {author} {\bibfnamefont {G.}~\bibnamefont
  {D'Amico}}, \bibinfo {author} {\bibfnamefont {P.}~\bibnamefont {Panci}},
  \bibinfo {author} {\bibfnamefont {A.}~\bibnamefont {Lupi}}, \bibinfo {author}
  {\bibfnamefont {S.}~\bibnamefont {Bovino}}, \ and\ \bibinfo {author}
  {\bibfnamefont {J.}~\bibnamefont {Silk}},\ }\href {\doibase
  10.1093/mnras/stx2419} {\bibfield  {journal} {\bibinfo  {journal} {Mon. Not.
  Roy. Astron. Soc.}\ }\textbf {\bibinfo {volume} {473}},\ \bibinfo {pages}
  {328} (\bibinfo {year} {2018})},\ \Eprint {http://arxiv.org/abs/1707.03419}
  {arXiv:1707.03419 [astro-ph.CO]} \BibitemShut {NoStop}%
\bibitem [{\citenamefont {{de Martino}}\ \emph {et~al.}(2020)\citenamefont {{de
  Martino}}, \citenamefont {{Chakrabarty}}, \citenamefont {{Cesare}},
  \citenamefont {{Gallo}}, \citenamefont {{Ostorero}},\ and\ \citenamefont
  {{Diaferio}}}]{2020Univ....6..107D}%
  \BibitemOpen
  \bibfield  {author} {\bibinfo {author} {\bibfnamefont {I.}~\bibnamefont {{de
  Martino}}}, \bibinfo {author} {\bibfnamefont {S.~S.}\ \bibnamefont
  {{Chakrabarty}}}, \bibinfo {author} {\bibfnamefont {V.}~\bibnamefont
  {{Cesare}}}, \bibinfo {author} {\bibfnamefont {A.}~\bibnamefont {{Gallo}}},
  \bibinfo {author} {\bibfnamefont {L.}~\bibnamefont {{Ostorero}}}, \ and\
  \bibinfo {author} {\bibfnamefont {A.}~\bibnamefont {{Diaferio}}},\ }\href
  {\doibase 10.3390/universe6080107} {\bibfield  {journal} {\bibinfo  {journal}
  {Universe}\ }\textbf {\bibinfo {volume} {6}},\ \bibinfo {pages} {107}
  (\bibinfo {year} {2020})},\ \Eprint {http://arxiv.org/abs/2007.15539}
  {arXiv:2007.15539 [astro-ph.CO]} \BibitemShut {NoStop}%
\bibitem [{\citenamefont {Abbott}\ \emph
  {et~al.}(2020{\natexlab{a}})\citenamefont {Abbott} \emph
  {et~al.}}]{Abbott:2020uma}%
  \BibitemOpen
  \bibfield  {author} {\bibinfo {author} {\bibfnamefont {B.}~\bibnamefont
  {Abbott}} \emph {et~al.} (\bibinfo {collaboration} {LIGO Scientific,
  Virgo}),\ }\href {\doibase 10.3847/2041-8213/ab75f5} {\bibfield  {journal}
  {\bibinfo  {journal} {Astrophys. J. Lett.}\ }\textbf {\bibinfo {volume}
  {892}},\ \bibinfo {pages} {L3} (\bibinfo {year} {2020}{\natexlab{a}})},\
  \Eprint {http://arxiv.org/abs/2001.01761} {arXiv:2001.01761 [astro-ph.HE]}
  \BibitemShut {NoStop}%
\bibitem [{\citenamefont {Kerr}(1963)}]{Kerr:1963ud}%
  \BibitemOpen
  \bibfield  {author} {\bibinfo {author} {\bibfnamefont {R.~P.}\ \bibnamefont
  {Kerr}},\ }\href {\doibase 10.1103/PhysRevLett.11.237} {\bibfield  {journal}
  {\bibinfo  {journal} {Phys. Rev. Lett.}\ }\textbf {\bibinfo {volume} {11}},\
  \bibinfo {pages} {237} (\bibinfo {year} {1963})}\BibitemShut {NoStop}%
\bibitem [{\citenamefont {Chandrasekhar}(1931)}]{Chandrasekhar:1931ih}%
  \BibitemOpen
  \bibfield  {author} {\bibinfo {author} {\bibfnamefont {S.}~\bibnamefont
  {Chandrasekhar}},\ }\href {\doibase 10.1086/143324} {\bibfield  {journal}
  {\bibinfo  {journal} {Astrophys. J.}\ }\textbf {\bibinfo {volume} {74}},\
  \bibinfo {pages} {81} (\bibinfo {year} {1931})}\BibitemShut {NoStop}%
\bibitem [{\citenamefont {Feng}\ \emph {et~al.}(2009)\citenamefont {Feng},
  \citenamefont {Kaplinghat}, \citenamefont {Tu},\ and\ \citenamefont
  {Yu}}]{Feng_2009}%
  \BibitemOpen
  \bibfield  {author} {\bibinfo {author} {\bibfnamefont {J.~L.}\ \bibnamefont
  {Feng}}, \bibinfo {author} {\bibfnamefont {M.}~\bibnamefont {Kaplinghat}},
  \bibinfo {author} {\bibfnamefont {H.}~\bibnamefont {Tu}}, \ and\ \bibinfo
  {author} {\bibfnamefont {H.-B.}\ \bibnamefont {Yu}},\ }\href {\doibase
  10.1088/1475-7516/2009/07/004} {\bibfield  {journal} {\bibinfo  {journal}
  {Journal of Cosmology and Astroparticle Physics}\ }\textbf {\bibinfo {volume}
  {2009}},\ \bibinfo {pages} {004} (\bibinfo {year} {2009})}\BibitemShut
  {NoStop}%
\bibitem [{\citenamefont {Buckley}\ and\ \citenamefont
  {DiFranzo}(2018)}]{PhysRevLett.120.051102}%
  \BibitemOpen
  \bibfield  {author} {\bibinfo {author} {\bibfnamefont {M.~R.}\ \bibnamefont
  {Buckley}}\ and\ \bibinfo {author} {\bibfnamefont {A.}~\bibnamefont
  {DiFranzo}},\ }\href {\doibase 10.1103/PhysRevLett.120.051102} {\bibfield
  {journal} {\bibinfo  {journal} {Phys. Rev. Lett.}\ }\textbf {\bibinfo
  {volume} {120}},\ \bibinfo {pages} {051102} (\bibinfo {year}
  {2018})}\BibitemShut {NoStop}%
\bibitem [{\citenamefont {Rosenberg}\ and\ \citenamefont
  {Fan}(2017)}]{PhysRevD.96.123001}%
  \BibitemOpen
  \bibfield  {author} {\bibinfo {author} {\bibfnamefont {E.}~\bibnamefont
  {Rosenberg}}\ and\ \bibinfo {author} {\bibfnamefont {J.}~\bibnamefont
  {Fan}},\ }\href {\doibase 10.1103/PhysRevD.96.123001} {\bibfield  {journal}
  {\bibinfo  {journal} {Phys. Rev. D}\ }\textbf {\bibinfo {volume} {96}},\
  \bibinfo {pages} {123001} (\bibinfo {year} {2017})}\BibitemShut {NoStop}%
\bibitem [{\citenamefont {Glover}(2012)}]{Glover2012}%
  \BibitemOpen
  \bibfield  {author} {\bibinfo {author} {\bibfnamefont {S.}~\bibnamefont
  {Glover}},\ }in\ \href {\doibase 10.1007/978-3-642-32362-1_3} {\emph
  {\bibinfo {booktitle} {The First Galaxies}}}\ (\bibinfo  {publisher}
  {Springer Berlin Heidelberg},\ \bibinfo {year} {2012})\ pp.\ \bibinfo {pages}
  {103--174}\BibitemShut {NoStop}%
\bibitem [{\citenamefont {{Rees}}(1976)}]{1976MNRAS.176..483R}%
  \BibitemOpen
  \bibfield  {author} {\bibinfo {author} {\bibfnamefont {M.~J.}\ \bibnamefont
  {{Rees}}},\ }\href {\doibase 10.1093/mnras/176.3.483} {\bibfield  {journal}
  {\bibinfo  {journal} {Monthly Notices of the Royal Astronomical Society}\
  }\textbf {\bibinfo {volume} {176}},\ \bibinfo {pages} {483} (\bibinfo {year}
  {1976})}\BibitemShut {NoStop}%
\bibitem [{\citenamefont {{Low}}\ and\ \citenamefont
  {{Lynden-Bell}}(1976)}]{1976MNRAS.176..367L}%
  \BibitemOpen
  \bibfield  {author} {\bibinfo {author} {\bibfnamefont {C.}~\bibnamefont
  {{Low}}}\ and\ \bibinfo {author} {\bibfnamefont {D.}~\bibnamefont
  {{Lynden-Bell}}},\ }\href {\doibase 10.1093/mnras/176.2.367} {\bibfield
  {journal} {\bibinfo  {journal} {Monthly Notices of the Royal Astronomical
  Society}\ }\textbf {\bibinfo {volume} {176}},\ \bibinfo {pages} {367}
  (\bibinfo {year} {1976})}\BibitemShut {NoStop}%
\bibitem [{\citenamefont {Abel}\ \emph {et~al.}(2002)\citenamefont {Abel},
  \citenamefont {Bryan},\ and\ \citenamefont {Norman}}]{Abel:2001pr}%
  \BibitemOpen
  \bibfield  {author} {\bibinfo {author} {\bibfnamefont {T.}~\bibnamefont
  {Abel}}, \bibinfo {author} {\bibfnamefont {G.~L.}\ \bibnamefont {Bryan}}, \
  and\ \bibinfo {author} {\bibfnamefont {M.~L.}\ \bibnamefont {Norman}},\
  }\href {\doibase 10.1126/science.1063991} {\bibfield  {journal} {\bibinfo
  {journal} {Science}\ }\textbf {\bibinfo {volume} {295}},\ \bibinfo {pages}
  {93} (\bibinfo {year} {2002})},\ \Eprint
  {http://arxiv.org/abs/astro-ph/0112088} {arXiv:astro-ph/0112088} \BibitemShut
  {NoStop}%
\bibitem [{\citenamefont {Bromm}\ and\ \citenamefont
  {Larson}(2004)}]{doi:10.1146/annurev.astro.42.053102.134034}%
  \BibitemOpen
  \bibfield  {author} {\bibinfo {author} {\bibfnamefont {V.}~\bibnamefont
  {Bromm}}\ and\ \bibinfo {author} {\bibfnamefont {R.~B.}\ \bibnamefont
  {Larson}},\ }\href {\doibase 10.1146/annurev.astro.42.053102.134034}
  {\bibfield  {journal} {\bibinfo  {journal} {Annual Review of Astronomy and
  Astrophysics}\ }\textbf {\bibinfo {volume} {42}},\ \bibinfo {pages} {79}
  (\bibinfo {year} {2004})},\ \Eprint
  {http://arxiv.org/abs/https://doi.org/10.1146/annurev.astro.42.053102.134034}
  {https://doi.org/10.1146/annurev.astro.42.053102.134034} \BibitemShut
  {NoStop}%
\bibitem [{\citenamefont {Shandera}\ \emph {et~al.}(2018)\citenamefont
  {Shandera}, \citenamefont {Jeong},\ and\ \citenamefont
  {Gebhardt}}]{Shandera:2018xkn}%
  \BibitemOpen
  \bibfield  {author} {\bibinfo {author} {\bibfnamefont {S.}~\bibnamefont
  {Shandera}}, \bibinfo {author} {\bibfnamefont {D.}~\bibnamefont {Jeong}}, \
  and\ \bibinfo {author} {\bibfnamefont {H.~S.~G.}\ \bibnamefont {Gebhardt}},\
  }\href {\doibase 10.1103/PhysRevLett.120.241102} {\bibfield  {journal}
  {\bibinfo  {journal} {Phys. Rev. Lett.}\ }\textbf {\bibinfo {volume} {120}},\
  \bibinfo {pages} {241102} (\bibinfo {year} {2018})},\ \Eprint
  {http://arxiv.org/abs/1802.08206} {arXiv:1802.08206 [astro-ph.CO]}
  \BibitemShut {NoStop}%
\bibitem [{\citenamefont {Peters}\ and\ \citenamefont
  {Mathews}(1963)}]{PhysRev.131.435}%
  \BibitemOpen
  \bibfield  {author} {\bibinfo {author} {\bibfnamefont {P.~C.}\ \bibnamefont
  {Peters}}\ and\ \bibinfo {author} {\bibfnamefont {J.}~\bibnamefont
  {Mathews}},\ }\href {\doibase 10.1103/PhysRev.131.435} {\bibfield  {journal}
  {\bibinfo  {journal} {Phys. Rev.}\ }\textbf {\bibinfo {volume} {131}},\
  \bibinfo {pages} {435} (\bibinfo {year} {1963})}\BibitemShut {NoStop}%
\bibitem [{\citenamefont {Hartwig}\ \emph {et~al.}(2016)\citenamefont
  {Hartwig}, \citenamefont {Volonteri}, \citenamefont {Bromm}, \citenamefont
  {Klessen}, \citenamefont {Barausse}, \citenamefont {Magg},\ and\
  \citenamefont {Stacy}}]{Hartwig:2016nde}%
  \BibitemOpen
  \bibfield  {author} {\bibinfo {author} {\bibfnamefont {T.}~\bibnamefont
  {Hartwig}}, \bibinfo {author} {\bibfnamefont {M.}~\bibnamefont {Volonteri}},
  \bibinfo {author} {\bibfnamefont {V.}~\bibnamefont {Bromm}}, \bibinfo
  {author} {\bibfnamefont {R.~S.}\ \bibnamefont {Klessen}}, \bibinfo {author}
  {\bibfnamefont {E.}~\bibnamefont {Barausse}}, \bibinfo {author}
  {\bibfnamefont {M.}~\bibnamefont {Magg}}, \ and\ \bibinfo {author}
  {\bibfnamefont {A.}~\bibnamefont {Stacy}},\ }\href {\doibase
  10.1093/mnrasl/slw074} {\bibfield  {journal} {\bibinfo  {journal} {Mon. Not.
  Roy. Astron. Soc.}\ }\textbf {\bibinfo {volume} {460}},\ \bibinfo {pages}
  {L74} (\bibinfo {year} {2016})},\ \Eprint {http://arxiv.org/abs/1603.05655}
  {arXiv:1603.05655 [astro-ph.GA]} \BibitemShut {NoStop}%
\bibitem [{\citenamefont {Abbott}\ \emph
  {et~al.}(2019{\natexlab{b}})\citenamefont {Abbott} \emph
  {et~al.}}]{LIGOScientific:2018mvr}%
  \BibitemOpen
  \bibfield  {author} {\bibinfo {author} {\bibfnamefont {B.}~\bibnamefont
  {Abbott}} \emph {et~al.} (\bibinfo {collaboration} {LIGO Scientific,
  Virgo}),\ }\href {\doibase 10.1103/PhysRevX.9.031040} {\bibfield  {journal}
  {\bibinfo  {journal} {Phys. Rev. X}\ }\textbf {\bibinfo {volume} {9}},\
  \bibinfo {pages} {031040} (\bibinfo {year} {2019}{\natexlab{b}})},\ \Eprint
  {http://arxiv.org/abs/1811.12907} {arXiv:1811.12907 [astro-ph.HE]}
  \BibitemShut {NoStop}%
\bibitem [{\citenamefont {Abadie}\ \emph {et~al.}(2010)\citenamefont {Abadie}
  \emph {et~al.}}]{Abadie:2010cg}%
  \BibitemOpen
  \bibfield  {author} {\bibinfo {author} {\bibfnamefont {J.}~\bibnamefont
  {Abadie}} \emph {et~al.} (\bibinfo {collaboration} {LIGO Scientific,
  VIRGO}),\ }\href@noop {} {\  (\bibinfo {year} {2010})},\ \Eprint
  {http://arxiv.org/abs/1003.2481} {arXiv:1003.2481 [gr-qc]} \BibitemShut
  {NoStop}%
\bibitem [{\citenamefont {Abbott}\ \emph
  {et~al.}(2020{\natexlab{b}})\citenamefont {Abbott} \emph
  {et~al.}}]{Abbott:2020niy}%
  \BibitemOpen
  \bibfield  {author} {\bibinfo {author} {\bibfnamefont {R.}~\bibnamefont
  {Abbott}} \emph {et~al.} (\bibinfo {collaboration} {LIGO Scientific,
  Virgo}),\ }\href@noop {} {\  (\bibinfo {year} {2020}{\natexlab{b}})},\
  \Eprint {http://arxiv.org/abs/2010.14527} {arXiv:2010.14527 [gr-qc]}
  \BibitemShut {NoStop}%
\bibitem [{\citenamefont {Abbott}\ \emph {et~al.}(2017)\citenamefont {Abbott}
  \emph {et~al.}}]{TheLIGOScientific:2017qsa}%
  \BibitemOpen
  \bibfield  {author} {\bibinfo {author} {\bibfnamefont {B.~P.}\ \bibnamefont
  {Abbott}} \emph {et~al.} (\bibinfo {collaboration} {Virgo, LIGO
  Scientific}),\ }\href {\doibase 10.1103/PhysRevLett.119.161101} {\bibfield
  {journal} {\bibinfo  {journal} {Phys. Rev. Lett.}\ }\textbf {\bibinfo
  {volume} {119}},\ \bibinfo {pages} {161101} (\bibinfo {year} {2017})},\
  \Eprint {http://arxiv.org/abs/1710.05832} {arXiv:1710.05832 [gr-qc]}
  \BibitemShut {NoStop}%
\bibitem [{\citenamefont {Biswas}\ \emph {et~al.}(2009)\citenamefont {Biswas},
  \citenamefont {Brady}, \citenamefont {Creighton},\ and\ \citenamefont
  {Fairhurst}}]{Biswas:2007ni}%
  \BibitemOpen
  \bibfield  {author} {\bibinfo {author} {\bibfnamefont {R.}~\bibnamefont
  {Biswas}}, \bibinfo {author} {\bibfnamefont {P.~R.}\ \bibnamefont {Brady}},
  \bibinfo {author} {\bibfnamefont {J.~D.~E.}\ \bibnamefont {Creighton}}, \
  and\ \bibinfo {author} {\bibfnamefont {S.}~\bibnamefont {Fairhurst}},\ }\href
  {\doibase 10.1088/0264-9381/26/17/175009, 10.1088/0264-9381/30/7/079502}
  {\bibfield  {journal} {\bibinfo  {journal} {Class. Quant. Grav.}\ }\textbf
  {\bibinfo {volume} {26}},\ \bibinfo {pages} {175009} (\bibinfo {year}
  {2009})},\ \bibinfo {note} {[Erratum: Class. Quant. Grav.30,079502(2013)]},\
  \Eprint {http://arxiv.org/abs/0710.0465} {arXiv:0710.0465 [gr-qc]}
  \BibitemShut {NoStop}%
\bibitem [{\citenamefont {Mukherjee}\ \emph {et~al.}(2018)\citenamefont
  {Mukherjee} \emph {et~al.}}]{Mukherjee:2018yra}%
  \BibitemOpen
  \bibfield  {author} {\bibinfo {author} {\bibfnamefont {D.}~\bibnamefont
  {Mukherjee}} \emph {et~al.},\ }\href@noop {} {\  (\bibinfo {year} {2018})},\
  \Eprint {http://arxiv.org/abs/1812.05121} {arXiv:1812.05121 [astro-ph.IM]}
  \BibitemShut {NoStop}%
\bibitem [{\citenamefont {{Farr}}\ \emph {et~al.}(2011)\citenamefont {{Farr}},
  \citenamefont {{Sravan}}, \citenamefont {{Cantrell}}, \citenamefont
  {{Kreidberg}}, \citenamefont {{Bailyn}}, \citenamefont {{Mandel}},\ and\
  \citenamefont {{Kalogera}}}]{2011ApJ...741..103F}%
  \BibitemOpen
  \bibfield  {author} {\bibinfo {author} {\bibfnamefont {W.~M.}\ \bibnamefont
  {{Farr}}}, \bibinfo {author} {\bibfnamefont {N.}~\bibnamefont {{Sravan}}},
  \bibinfo {author} {\bibfnamefont {A.}~\bibnamefont {{Cantrell}}}, \bibinfo
  {author} {\bibfnamefont {L.}~\bibnamefont {{Kreidberg}}}, \bibinfo {author}
  {\bibfnamefont {C.~D.}\ \bibnamefont {{Bailyn}}}, \bibinfo {author}
  {\bibfnamefont {I.}~\bibnamefont {{Mandel}}}, \ and\ \bibinfo {author}
  {\bibfnamefont {V.}~\bibnamefont {{Kalogera}}},\ }\href {\doibase
  10.1088/0004-637X/741/2/103} {\bibfield  {journal} {\bibinfo  {journal}
  {\apj}\ }\textbf {\bibinfo {volume} {741}},\ \bibinfo {eid} {103} (\bibinfo
  {year} {2011})},\ \Eprint {http://arxiv.org/abs/1011.1459} {arXiv:1011.1459}
  \BibitemShut {NoStop}%
\bibitem [{\citenamefont {Faber}\ and\ \citenamefont
  {Rasio}(2012)}]{Faber:2012rw}%
  \BibitemOpen
  \bibfield  {author} {\bibinfo {author} {\bibfnamefont {J.~A.}\ \bibnamefont
  {Faber}}\ and\ \bibinfo {author} {\bibfnamefont {F.~A.}\ \bibnamefont
  {Rasio}},\ }\href {\doibase 10.12942/lrr-2012-8} {\bibfield  {journal}
  {\bibinfo  {journal} {Living Rev. Rel.}\ }\textbf {\bibinfo {volume} {15}},\
  \bibinfo {pages} {8} (\bibinfo {year} {2012})},\ \Eprint
  {http://arxiv.org/abs/1204.3858} {arXiv:1204.3858 [gr-qc]} \BibitemShut
  {NoStop}%
\bibitem [{\citenamefont {Gupta}\ \emph {et~al.}(2020)\citenamefont {Gupta},
  \citenamefont {Gerosa}, \citenamefont {Arun}, \citenamefont {Berti},
  \citenamefont {Farr},\ and\ \citenamefont {Sathyaprakash}}]{Gupta:2019nwj}%
  \BibitemOpen
  \bibfield  {author} {\bibinfo {author} {\bibfnamefont {A.}~\bibnamefont
  {Gupta}}, \bibinfo {author} {\bibfnamefont {D.}~\bibnamefont {Gerosa}},
  \bibinfo {author} {\bibfnamefont {K.}~\bibnamefont {Arun}}, \bibinfo {author}
  {\bibfnamefont {E.}~\bibnamefont {Berti}}, \bibinfo {author} {\bibfnamefont
  {W.~M.}\ \bibnamefont {Farr}}, \ and\ \bibinfo {author} {\bibfnamefont
  {B.}~\bibnamefont {Sathyaprakash}},\ }\href {\doibase
  10.1103/PhysRevD.101.103036} {\bibfield  {journal} {\bibinfo  {journal}
  {Phys. Rev. D}\ }\textbf {\bibinfo {volume} {101}},\ \bibinfo {pages}
  {103036} (\bibinfo {year} {2020})},\ \Eprint
  {http://arxiv.org/abs/1909.05804} {arXiv:1909.05804 [gr-qc]} \BibitemShut
  {NoStop}%
\bibitem [{\citenamefont {Bramante}\ \emph {et~al.}(2018)\citenamefont
  {Bramante}, \citenamefont {Linden},\ and\ \citenamefont
  {Tsai}}]{Bramante:2017ulk}%
  \BibitemOpen
  \bibfield  {author} {\bibinfo {author} {\bibfnamefont {J.}~\bibnamefont
  {Bramante}}, \bibinfo {author} {\bibfnamefont {T.}~\bibnamefont {Linden}}, \
  and\ \bibinfo {author} {\bibfnamefont {Y.-D.}\ \bibnamefont {Tsai}},\ }\href
  {\doibase 10.1103/PhysRevD.97.055016} {\bibfield  {journal} {\bibinfo
  {journal} {Phys. Rev. D}\ }\textbf {\bibinfo {volume} {97}},\ \bibinfo
  {pages} {055016} (\bibinfo {year} {2018})},\ \Eprint
  {http://arxiv.org/abs/1706.00001} {arXiv:1706.00001 [hep-ph]} \BibitemShut
  {NoStop}%
\bibitem [{\citenamefont {Belczynski}\ \emph {et~al.}(2012)\citenamefont
  {Belczynski}, \citenamefont {Wiktorowicz}, \citenamefont {Fryer},
  \citenamefont {Holz},\ and\ \citenamefont {Kalogera}}]{Belczynski_2012}%
  \BibitemOpen
  \bibfield  {author} {\bibinfo {author} {\bibfnamefont {K.}~\bibnamefont
  {Belczynski}}, \bibinfo {author} {\bibfnamefont {G.}~\bibnamefont
  {Wiktorowicz}}, \bibinfo {author} {\bibfnamefont {C.~L.}\ \bibnamefont
  {Fryer}}, \bibinfo {author} {\bibfnamefont {D.~E.}\ \bibnamefont {Holz}}, \
  and\ \bibinfo {author} {\bibfnamefont {V.}~\bibnamefont {Kalogera}},\ }\href
  {\doibase 10.1088/0004-637x/757/1/91} {\bibfield  {journal} {\bibinfo
  {journal} {The Astrophysical Journal}\ }\textbf {\bibinfo {volume} {757}},\
  \bibinfo {pages} {91} (\bibinfo {year} {2012})}\BibitemShut {NoStop}%
\bibitem [{\citenamefont {Abbott}\ \emph
  {et~al.}(2020{\natexlab{c}})\citenamefont {Abbott} \emph
  {et~al.}}]{Abbott:2020khf}%
  \BibitemOpen
  \bibfield  {author} {\bibinfo {author} {\bibfnamefont {R.}~\bibnamefont
  {Abbott}} \emph {et~al.} (\bibinfo {collaboration} {LIGO Scientific,
  Virgo}),\ }\href {\doibase 10.3847/2041-8213/ab960f} {\bibfield  {journal}
  {\bibinfo  {journal} {Astrophys. J. Lett.}\ }\textbf {\bibinfo {volume}
  {896}},\ \bibinfo {pages} {L44} (\bibinfo {year} {2020}{\natexlab{c}})},\
  \Eprint {http://arxiv.org/abs/2006.12611} {arXiv:2006.12611 [astro-ph.HE]}
  \BibitemShut {NoStop}%
\bibitem [{\citenamefont {Markevitch}\ \emph {et~al.}(2004)\citenamefont
  {Markevitch}, \citenamefont {Gonzalez}, \citenamefont {Clowe}, \citenamefont
  {Vikhlinin}, \citenamefont {David}, \citenamefont {Forman}, \citenamefont
  {Jones}, \citenamefont {Murray},\ and\ \citenamefont
  {Tucker}}]{Markevitch:2003at}%
  \BibitemOpen
  \bibfield  {author} {\bibinfo {author} {\bibfnamefont {M.}~\bibnamefont
  {Markevitch}}, \bibinfo {author} {\bibfnamefont {A.}~\bibnamefont
  {Gonzalez}}, \bibinfo {author} {\bibfnamefont {D.}~\bibnamefont {Clowe}},
  \bibinfo {author} {\bibfnamefont {A.}~\bibnamefont {Vikhlinin}}, \bibinfo
  {author} {\bibfnamefont {L.}~\bibnamefont {David}}, \bibinfo {author}
  {\bibfnamefont {W.}~\bibnamefont {Forman}}, \bibinfo {author} {\bibfnamefont
  {C.}~\bibnamefont {Jones}}, \bibinfo {author} {\bibfnamefont
  {S.}~\bibnamefont {Murray}}, \ and\ \bibinfo {author} {\bibfnamefont
  {W.}~\bibnamefont {Tucker}},\ }\href {\doibase 10.1086/383178} {\bibfield
  {journal} {\bibinfo  {journal} {Astrophys. J.}\ }\textbf {\bibinfo {volume}
  {606}},\ \bibinfo {pages} {819} (\bibinfo {year} {2004})},\ \Eprint
  {http://arxiv.org/abs/astro-ph/0309303} {arXiv:astro-ph/0309303} \BibitemShut
  {NoStop}%
\bibitem [{\citenamefont {Herrmann}\ \emph {et~al.}(2016)\citenamefont
  {Herrmann}, \citenamefont {Hunter},\ and\ \citenamefont
  {Elmegreen}}]{Herrmann_2016}%
  \BibitemOpen
  \bibfield  {author} {\bibinfo {author} {\bibfnamefont {K.~A.}\ \bibnamefont
  {Herrmann}}, \bibinfo {author} {\bibfnamefont {D.~A.}\ \bibnamefont
  {Hunter}}, \ and\ \bibinfo {author} {\bibfnamefont {B.~G.}\ \bibnamefont
  {Elmegreen}},\ }\href {\doibase 10.3847/0004-6256/151/6/145} {\bibfield
  {journal} {\bibinfo  {journal} {The Astronomical Journal}\ }\textbf {\bibinfo
  {volume} {151}},\ \bibinfo {pages} {145} (\bibinfo {year}
  {2016})}\BibitemShut {NoStop}%
\bibitem [{\citenamefont {McConnachie}(2012)}]{McConnachie_2012}%
  \BibitemOpen
  \bibfield  {author} {\bibinfo {author} {\bibfnamefont {A.~W.}\ \bibnamefont
  {McConnachie}},\ }\href {\doibase 10.1088/0004-6256/144/1/4} {\bibfield
  {journal} {\bibinfo  {journal} {The Astronomical Journal}\ }\textbf {\bibinfo
  {volume} {144}},\ \bibinfo {pages} {4} (\bibinfo {year} {2012})}\BibitemShut
  {NoStop}%
\bibitem [{\citenamefont {Hawking}(1971)}]{Hawking1971}%
  \BibitemOpen
  \bibfield  {author} {\bibinfo {author} {\bibfnamefont {S.}~\bibnamefont
  {Hawking}},\ }\href {\doibase 10.1093/mnras/152.1.75} {\bibfield  {journal}
  {\bibinfo  {journal} {Monthly Notices of the Royal Astronomical Society}\
  }\textbf {\bibinfo {volume} {152}},\ \bibinfo {pages} {75} (\bibinfo {year}
  {1971})}\BibitemShut {NoStop}%
\bibitem [{\citenamefont {Chapline}(1975)}]{Chapline1975}%
  \BibitemOpen
  \bibfield  {author} {\bibinfo {author} {\bibfnamefont {G.~F.}\ \bibnamefont
  {Chapline}},\ }\href {\doibase 10.1038/253251a0} {\bibfield  {journal}
  {\bibinfo  {journal} {Nature}\ }\textbf {\bibinfo {volume} {253}},\ \bibinfo
  {pages} {251} (\bibinfo {year} {1975})}\BibitemShut {NoStop}%
\bibitem [{\citenamefont {Alcock}\ \emph {et~al.}(1993)\citenamefont {Alcock}
  \emph {et~al.}}]{Alcock:1993eu}%
  \BibitemOpen
  \bibfield  {author} {\bibinfo {author} {\bibfnamefont {C.}~\bibnamefont
  {Alcock}} \emph {et~al.} (\bibinfo {collaboration} {Supernova Cosmology
  Project}),\ }\href {\doibase 10.1038/365621a0} {\bibfield  {journal}
  {\bibinfo  {journal} {Nature}\ }\textbf {\bibinfo {volume} {365}},\ \bibinfo
  {pages} {621} (\bibinfo {year} {1993})},\ \Eprint
  {http://arxiv.org/abs/astro-ph/9309052} {arXiv:astro-ph/9309052} \BibitemShut
  {NoStop}%
\bibitem [{\citenamefont {Nakamura}\ \emph {et~al.}(1997)\citenamefont
  {Nakamura}, \citenamefont {Sasaki}, \citenamefont {Tanaka},\ and\
  \citenamefont {Thorne}}]{Nakamura:1997sm}%
  \BibitemOpen
  \bibfield  {author} {\bibinfo {author} {\bibfnamefont {T.}~\bibnamefont
  {Nakamura}}, \bibinfo {author} {\bibfnamefont {M.}~\bibnamefont {Sasaki}},
  \bibinfo {author} {\bibfnamefont {T.}~\bibnamefont {Tanaka}}, \ and\ \bibinfo
  {author} {\bibfnamefont {K.~S.}\ \bibnamefont {Thorne}},\ }\href {\doibase
  10.1086/310886} {\bibfield  {journal} {\bibinfo  {journal} {Astrophys. J.}\
  }\textbf {\bibinfo {volume} {487}},\ \bibinfo {pages} {L139} (\bibinfo {year}
  {1997})},\ \Eprint {http://arxiv.org/abs/astro-ph/9708060}
  {arXiv:astro-ph/9708060 [astro-ph]} \BibitemShut {NoStop}%
\bibitem [{\citenamefont {Allsman}\ \emph {et~al.}(2001)\citenamefont {Allsman}
  \emph {et~al.}}]{Allsman:2000kg}%
  \BibitemOpen
  \bibfield  {author} {\bibinfo {author} {\bibfnamefont {R.~A.}\ \bibnamefont
  {Allsman}} \emph {et~al.} (\bibinfo {collaboration} {Macho}),\ }\href
  {\doibase 10.1086/319636} {\bibfield  {journal} {\bibinfo  {journal}
  {Astrophys. J.}\ }\textbf {\bibinfo {volume} {550}},\ \bibinfo {pages} {L169}
  (\bibinfo {year} {2001})},\ \Eprint {http://arxiv.org/abs/astro-ph/0011506}
  {arXiv:astro-ph/0011506 [astro-ph]} \BibitemShut {NoStop}%
\bibitem [{\citenamefont {Tisserand}\ \emph {et~al.}(2007)\citenamefont
  {Tisserand} \emph {et~al.}}]{Tisserand:2006zx}%
  \BibitemOpen
  \bibfield  {author} {\bibinfo {author} {\bibfnamefont {P.}~\bibnamefont
  {Tisserand}} \emph {et~al.} (\bibinfo {collaboration} {EROS-2}),\ }\href
  {\doibase 10.1051/0004-6361:20066017} {\bibfield  {journal} {\bibinfo
  {journal} {Astron. Astrophys.}\ }\textbf {\bibinfo {volume} {469}},\ \bibinfo
  {pages} {387} (\bibinfo {year} {2007})},\ \Eprint
  {http://arxiv.org/abs/astro-ph/0607207} {arXiv:astro-ph/0607207 [astro-ph]}
  \BibitemShut {NoStop}%
\bibitem [{\citenamefont {Abbott}\ \emph {et~al.}(2008)\citenamefont {Abbott}
  \emph {et~al.}}]{Abbott:2007xi}%
  \BibitemOpen
  \bibfield  {author} {\bibinfo {author} {\bibfnamefont {B.}~\bibnamefont
  {Abbott}} \emph {et~al.} (\bibinfo {collaboration} {LIGO Scientific}),\
  }\href {\doibase 10.1103/PhysRevD.77.062002} {\bibfield  {journal} {\bibinfo
  {journal} {Phys. Rev.}\ }\textbf {\bibinfo {volume} {D77}},\ \bibinfo {pages}
  {062002} (\bibinfo {year} {2008})},\ \Eprint {http://arxiv.org/abs/0704.3368}
  {arXiv:0704.3368 [gr-qc]} \BibitemShut {NoStop}%
\bibitem [{\citenamefont {Wade}(2015)}]{wade2015}%
  \BibitemOpen
  \bibfield  {author} {\bibinfo {author} {\bibfnamefont {M.}~\bibnamefont
  {Wade}},\ }\emph {\bibinfo {title} {{Gravitational-wave Science with the
  Laser Interferometer Gravitational-wave Observatory}}},\ \href
  {https://dcc.ligo.org/LIGO-P1500068/public} {Ph.D. thesis},\ \bibinfo
  {school} {University of Wisconsin-Milwaukee} (\bibinfo {year} {2015}),\
  \Eprint {http://arxiv.org/abs/LIGO-P1500068} {LIGO-P1500068} \BibitemShut
  {NoStop}%
\bibitem [{\citenamefont {Brandt}(2016)}]{Brandt:2016aco}%
  \BibitemOpen
  \bibfield  {author} {\bibinfo {author} {\bibfnamefont {T.~D.}\ \bibnamefont
  {Brandt}},\ }\href {\doibase 10.3847/2041-8205/824/2/L31} {\bibfield
  {journal} {\bibinfo  {journal} {Astrophys. J.}\ }\textbf {\bibinfo {volume}
  {824}},\ \bibinfo {pages} {L31} (\bibinfo {year} {2016})},\ \Eprint
  {http://arxiv.org/abs/1605.03665} {arXiv:1605.03665 [astro-ph.GA]}
  \BibitemShut {NoStop}%
\bibitem [{\citenamefont {Koushiappas}\ and\ \citenamefont
  {Loeb}(2017)}]{Koushiappas:2017chw}%
  \BibitemOpen
  \bibfield  {author} {\bibinfo {author} {\bibfnamefont {S.~M.}\ \bibnamefont
  {Koushiappas}}\ and\ \bibinfo {author} {\bibfnamefont {A.}~\bibnamefont
  {Loeb}},\ }\href {\doibase 10.1103/PhysRevLett.119.041102} {\bibfield
  {journal} {\bibinfo  {journal} {Phys. Rev. Lett.}\ }\textbf {\bibinfo
  {volume} {119}},\ \bibinfo {pages} {041102} (\bibinfo {year} {2017})},\
  \Eprint {http://arxiv.org/abs/1704.01668} {arXiv:1704.01668 [astro-ph.GA]}
  \BibitemShut {NoStop}%
\bibitem [{\citenamefont {{Wyrzykowski}}\ \emph {et~al.}(2011)\citenamefont
  {{Wyrzykowski}}, \citenamefont {{Skowron}}, \citenamefont {{Koz{\l}owski}},
  \citenamefont {{Udalski}}, \citenamefont {{Szyma{\'n}ski}}, \citenamefont
  {{Kubiak}}, \citenamefont {{Pietrzy{\'n}ski}}, \citenamefont
  {{Soszy{\'n}ski}}, \citenamefont {{Szewczyk}}, \citenamefont {{Ulaczyk}},
  \citenamefont {{Poleski}},\ and\ \citenamefont {{Tisserand}}}]{ogle}%
  \BibitemOpen
  \bibfield  {author} {\bibinfo {author} {\bibfnamefont {L.}~\bibnamefont
  {{Wyrzykowski}}}, \bibinfo {author} {\bibfnamefont {J.}~\bibnamefont
  {{Skowron}}}, \bibinfo {author} {\bibfnamefont {S.}~\bibnamefont
  {{Koz{\l}owski}}}, \bibinfo {author} {\bibfnamefont {A.}~\bibnamefont
  {{Udalski}}}, \bibinfo {author} {\bibfnamefont {M.~K.}\ \bibnamefont
  {{Szyma{\'n}ski}}}, \bibinfo {author} {\bibfnamefont {M.}~\bibnamefont
  {{Kubiak}}}, \bibinfo {author} {\bibfnamefont {G.}~\bibnamefont
  {{Pietrzy{\'n}ski}}}, \bibinfo {author} {\bibfnamefont {I.}~\bibnamefont
  {{Soszy{\'n}ski}}}, \bibinfo {author} {\bibfnamefont {O.}~\bibnamefont
  {{Szewczyk}}}, \bibinfo {author} {\bibfnamefont {K.}~\bibnamefont
  {{Ulaczyk}}}, \bibinfo {author} {\bibfnamefont {R.}~\bibnamefont
  {{Poleski}}}, \ and\ \bibinfo {author} {\bibfnamefont {P.}~\bibnamefont
  {{Tisserand}}},\ }\href {\doibase 10.1111/j.1365-2966.2011.19243.x}
  {\bibfield  {journal} {\bibinfo  {journal} {Monthly Notices of the Royal
  Astronomical Society}\ }\textbf {\bibinfo {volume} {416}},\ \bibinfo {pages}
  {2949} (\bibinfo {year} {2011})},\ \Eprint {http://arxiv.org/abs/1106.2925}
  {arXiv:1106.2925 [astro-ph.GA]} \BibitemShut {NoStop}%
\bibitem [{\citenamefont {Gaggero}\ \emph {et~al.}(2017)\citenamefont
  {Gaggero}, \citenamefont {Bertone}, \citenamefont {Calore}, \citenamefont
  {Connors}, \citenamefont {Lovell}, \citenamefont {Markoff},\ and\
  \citenamefont {Storm}}]{Gaggero:2016dpq}%
  \BibitemOpen
  \bibfield  {author} {\bibinfo {author} {\bibfnamefont {D.}~\bibnamefont
  {Gaggero}}, \bibinfo {author} {\bibfnamefont {G.}~\bibnamefont {Bertone}},
  \bibinfo {author} {\bibfnamefont {F.}~\bibnamefont {Calore}}, \bibinfo
  {author} {\bibfnamefont {R.~M.~T.}\ \bibnamefont {Connors}}, \bibinfo
  {author} {\bibfnamefont {M.}~\bibnamefont {Lovell}}, \bibinfo {author}
  {\bibfnamefont {S.}~\bibnamefont {Markoff}}, \ and\ \bibinfo {author}
  {\bibfnamefont {E.}~\bibnamefont {Storm}},\ }\href {\doibase
  10.1103/PhysRevLett.118.241101} {\bibfield  {journal} {\bibinfo  {journal}
  {Phys. Rev. Lett.}\ }\textbf {\bibinfo {volume} {118}},\ \bibinfo {pages}
  {241101} (\bibinfo {year} {2017})},\ \Eprint
  {http://arxiv.org/abs/1612.00457} {arXiv:1612.00457 [astro-ph.HE]}
  \BibitemShut {NoStop}%
\bibitem [{\citenamefont {Raidal}\ \emph {et~al.}(2017)\citenamefont {Raidal},
  \citenamefont {Vaskonen},\ and\ \citenamefont {Veerm\"ae}}]{Raidal:2017mfl}%
  \BibitemOpen
  \bibfield  {author} {\bibinfo {author} {\bibfnamefont {M.}~\bibnamefont
  {Raidal}}, \bibinfo {author} {\bibfnamefont {V.}~\bibnamefont {Vaskonen}}, \
  and\ \bibinfo {author} {\bibfnamefont {H.}~\bibnamefont {Veerm\"ae}},\ }\href
  {\doibase 10.1088/1475-7516/2017/09/037} {\bibfield  {journal} {\bibinfo
  {journal} {JCAP}\ }\textbf {\bibinfo {volume} {09}},\ \bibinfo {pages} {037}
  (\bibinfo {year} {2017})},\ \Eprint {http://arxiv.org/abs/1707.01480}
  {arXiv:1707.01480 [astro-ph.CO]} \BibitemShut {NoStop}%
\bibitem [{\citenamefont {Raidal}\ \emph {et~al.}(2019)\citenamefont {Raidal},
  \citenamefont {Spethmann}, \citenamefont {Vaskonen},\ and\ \citenamefont
  {Veerm\"ae}}]{Raidal:2018bbj}%
  \BibitemOpen
  \bibfield  {author} {\bibinfo {author} {\bibfnamefont {M.}~\bibnamefont
  {Raidal}}, \bibinfo {author} {\bibfnamefont {C.}~\bibnamefont {Spethmann}},
  \bibinfo {author} {\bibfnamefont {V.}~\bibnamefont {Vaskonen}}, \ and\
  \bibinfo {author} {\bibfnamefont {H.}~\bibnamefont {Veerm\"ae}},\ }\href
  {\doibase 10.1088/1475-7516/2019/02/018} {\bibfield  {journal} {\bibinfo
  {journal} {JCAP}\ }\textbf {\bibinfo {volume} {02}},\ \bibinfo {pages} {018}
  (\bibinfo {year} {2019})},\ \Eprint {http://arxiv.org/abs/1812.01930}
  {arXiv:1812.01930 [astro-ph.CO]} \BibitemShut {NoStop}%
\bibitem [{\citenamefont {Ali-Ha{\"\i}moud}\ \emph {et~al.}(2017)\citenamefont
  {Ali-Ha{\"\i}moud}, \citenamefont {Kovetz},\ and\ \citenamefont
  {Kamionkowski}}]{Ali-Haimoud:2017rtz}%
  \BibitemOpen
  \bibfield  {author} {\bibinfo {author} {\bibfnamefont {Y.}~\bibnamefont
  {Ali-Ha{\"\i}moud}}, \bibinfo {author} {\bibfnamefont {E.~D.}\ \bibnamefont
  {Kovetz}}, \ and\ \bibinfo {author} {\bibfnamefont {M.}~\bibnamefont
  {Kamionkowski}},\ }\href {\doibase 10.1103/PhysRevD.96.123523} {\bibfield
  {journal} {\bibinfo  {journal} {Phys. Rev.}\ }\textbf {\bibinfo {volume}
  {D96}},\ \bibinfo {pages} {123523} (\bibinfo {year} {2017})},\ \Eprint
  {http://arxiv.org/abs/1709.06576} {arXiv:1709.06576 [astro-ph.CO]}
  \BibitemShut {NoStop}%
\bibitem [{\citenamefont {Abbott}\ \emph
  {et~al.}(2019{\natexlab{c}})\citenamefont {Abbott} \emph
  {et~al.}}]{Authors:2019qbw}%
  \BibitemOpen
  \bibfield  {author} {\bibinfo {author} {\bibfnamefont {B.~P.}\ \bibnamefont
  {Abbott}} \emph {et~al.} (\bibinfo {collaboration} {LIGO Scientific,
  Virgo}),\ }\href {\doibase 10.1103/PhysRevLett.123.161102} {\bibfield
  {journal} {\bibinfo  {journal} {Phys. Rev. Lett.}\ }\textbf {\bibinfo
  {volume} {123}},\ \bibinfo {pages} {161102} (\bibinfo {year}
  {2019}{\natexlab{c}})},\ \Eprint {http://arxiv.org/abs/1904.08976}
  {arXiv:1904.08976 [astro-ph.CO]} \BibitemShut {NoStop}%
\end{thebibliography}%

\end{document}